%% file: Gravity-Edge-Modes-1.tex
\documentclass[11pt,showlabels]{article}

\usepackage{authblk}
\usepackage{amsmath}
\usepackage{amsfonts}
\usepackage{amssymb}
\usepackage{mathrsfs}
\usepackage{slashed}
\usepackage{color}
\usepackage{mathtools}
\usepackage{marvosym}
\usepackage{amssymb}
\usepackage{dsfont}
\usepackage{slashed}
\usepackage{fullpage}
\usepackage{cite}
\usepackage{accents}
\usepackage[colorlinks=true,linkcolor=blue,citecolor=magenta,linktocpage=true]{hyperref}
\usepackage{titlesec}
\titleformat*{\section}{\normalsize\bfseries}
\titleformat*{\subsection}{\normalsize\bfseries}
\titleformat*{\subsubsection}{\normalsize\bfseries}
\usepackage{xcolor}
\usepackage{subcaption}
\usepackage{url}
\DeclareMathAlphabet{\bbvar}{U}{BOONDOX-ds}{m}{n}

\newtheorem{dfn}{Definition}[section]

\newtheorem{prop}{Proposition}[section]

\makeatletter
\renewcommand{\@dotsep}{10000}
\makeatother

\input Macro.tex
\begin{document}

\title{\Large{\textbf{\sffamily Edge modes of gravity - I:\\Corner potentials and charges}}}
\author{\sffamily Laurent Freidel$^1$, Marc Geiller$^2$, Daniele Pranzetti$^1$}
\date{\small{\textit{
$^1$Perimeter Institute for Theoretical Physics,\\ 31 Caroline Street North, Waterloo, Ontario, Canada N2L 2Y5\\
$^2$Univ Lyon, ENS de Lyon, Univ Claude Bernard Lyon 1,\\ CNRS, Laboratoire de Physique, UMR 5672, F-69342 Lyon, France\\}}}

\maketitle

\begin{abstract}

This is the first paper in a series devoted to understanding the classical and quantum nature of edge modes and symmetries in gravitational systems. The goal of this analysis is to: i) achieve a clear understanding of how different formulations of gravity provide non-trivial representations of different sectors of the corner symmetry algebra, and ii) set the foundations of a new proposal for states of quantum geometry as representation states of this corner symmetry algebra. In this first paper we explain how different formulations of gravity, in both metric and tetrad variables, share the same bulk symplectic structure but differ at the corner, and in turn lead to inequivalent representations of the corner symmetry algebra. This provides an organizing criterion for formulations of gravity depending on how big  the physical symmetry group that is non-trivially represented at the corner is. This principle can be used as  a ``treasure map'' revealing new clues and routes in the quest for quantum gravity. Building up on these results, we perform a detailed analysis of the corner pre-symplectic potential and symmetries of Einstein--Cartan--Holst gravity in \cite{Edge-Mode-II}, use this to provide a new look at the simplicity constraints in \cite{Edge-Mode-III}, and tackle the quantization in \cite{Edge-Mode-IV}.

\end{abstract}

\thispagestyle{empty}
\newpage
\setcounter{page}{1}

\hrule
\tableofcontents
\vspace{0.7cm}
\hrule

\newpage

\section{Introduction}
\label{sec:1}

What are the symmetries of gravity? This is the question which we would like to properly pose and answer in the present series of articles, with the viewpoint that doing so is necessary in order to address the problem of quantum gravity. Gravity being a gauge theory, its invariance under spacetime diffeomorphisms only represents a \textit{gauge} symmetry, and not a \textit{physical} symmetry. Gauge symmetries only label gauge redundancies and have a vanishing charge. As such, they cannot be used to label or distinguish physical states of a theory (of, say, quantum gravity), since by definition their Hamiltonian generators vanish on such physical states. However, this situation changes in the presence of boundaries (be they asymptotic or at finite distance).

When considering bounded regions, a subset of transformations, which are gauge in the bulk, become physical symmetries on the boundary and acquire a non-vanishing charge. The fact that important physics is unfolding at the boundaries of gauge theories has been recognized early on, prominently in condensed matter systems \cite{PhysRevLett.62.82,Frohlich:1990xz,Balachandran:1991dw,Wen:1992vi} and in the context of black holes \cite{ Carlip:1994gy, Balachandran:1994vi, Balachandran:1995qa, Ashtekar:1997yu, Banados:1998ta, Carlip:2005zn, Engle:2010kt, Ghosh:2014rra}, while a few prescient works assigned non-trivial degrees of freedom to general surfaces \cite{Husain:1998zw,Donnelly:2008vx,Bodendorfer:2014fua}, relating them to the concept of entanglement entropy. The literature has assigned many names to the degrees of freedom involved in this boundary physics, including edge states, edge modes, boundary degrees of freedom, and would-be-gauge degrees of freedom. The most notable feature, which is the focus of this series of articles, is that the boundary charges of \textit{physical} symmetries, which are located on codimension-2 spheres (or corners), posses a non-trivial algebra, and that this latter is typically vastly different from the algebra of \textit{gauge} symmetries\footnote{The algebra of corner symmetries is for example often a current algebra with possible central extensions, which needs to be represented non-trivially at the quantum level. On the other hand, and by definition, the algebra of gauge transformations should be anomaly-free (i.e. without central extensions), and they are trivially represented at the quantum level.}. Our goal is to understand, in a systematic manner, the nature of this corner symmetry algebra in the case of gravity, and to use this as a guiding principle for quantum gravity.

Throughout the years, a very substantial amount of work has been dedicated to the study of these corner charges, their algebra, and their possible physical applications \cite{Brown:1986nw,Coussaert:1995zp,Barnich:2001jy,Barnich:2007bf,Comp_re_2008,Compere:2014cna,Donnay:2016ejv,Oblak:2016eij,Hawking:2016msc,Afshar:2017okz,Carlip:2017xne,Afshar:2019axx}. This has lead to a zoology of boundary symmetry algebras depending (for a given theory) on the location and the type of boundary, and on the boundary conditions being imposed. Taking the viewpoint that representations of a symmetry algebra provide an organizing principle for states of a quantum theory, one would like to find the most general boundary symmetry algebra, which would allow in turn to understand its reduction to the various subalgebras which have been discovered in the literature. This has motivated work on the study of the most general boundary conditions in e.g. 3-dimensional gravity \cite{Grumiller:2016pqb,Grumiller:2017sjh}. Conveniently, there is a level at which one can discuss the boundary symmetries independently of a choice of boundary conditions. This will enable us to properly frame the question raised above: What are the symmetries of gravity?

The central importance of symmetries stems from the fact that they give us a firm non-perturbative handle on quantization, even in the context where the quantum theory is not known such as in gravity. One of the main reasons behind this can be understood in terms of the Kirillov orbit method \cite{Kirillov}. This method, which is available when a classical symmetry is acting on a physical system, allows to pull-back purely quantum notions into the classical realm, thereby rendering the gap between quantum and classical extremely thin. This is a framework which associates to classical symmetries and their canonical action a notion of \emph{representations} (as labels of the classical coadjoint orbits), of weights (as Casimirs of the Poisson algebra of charges), of states (as Lagrangian leafs of the symplectic orbits), and of characters (as Fourier transformation of the orbit measure). Even if the Kirillov correspondence is not rigorously proven for the infinite-dimensional symmetry groups which we consider here, it is known to hold true for a large class of compact, non-compact \cite{Rossman} and even infinite-dimensional groups \cite{Frenkel1984OrbitalTF}. We will use it as a ``treasure map'' to guide us into a pre-quantization program for quantum gravity. We will exploit in particular the central concept of ``representation'' for the classical symmetry group in the Kirillov sense. As we will argue, this new concept of representation associated with a gravitational symmetry group provides us with an invaluable tool to grasp some key and universal elements of the elusive quantum theory of gravity. Some of the key aspects pertaining to this have already been explored in \cite{Freidel:2015gpa,Freidel:2016bxd,Freidel:2019ees,Freidel:2019ofr}.

The symmetry content of a gauge theory is best elucidated in the covariant phase space formalism \cite{Kijowski1976ACS, Gawdzki1991ClassicalOO, Crnkovic:1986ex,Ashtekar:1990gc,Lee:1990nz,Wald:1999wa}, which we  therefore adopt. The most minimal setup in which physical symmetries and their charges appear is when considering an entangling wedge. This is a foliation of the spacetime manifold $M$ into Cauchy hypersurfaces $\Sigma$ which all meet at a codimension-2 corner\footnote{$S$ is the boundary of $\Sigma$, and it is also a \emph{corner} of spacetime. We use the name corner for $S$ to insist on the fact that it is a codimension-2 surface, and to distinguish it from spacetime boundaries which are codimension-1 surfaces.} $S$. 
The local geometry of this entangling wedge is represented on Figure \ref{Wedge}. Our goal is to explain, in gravity, what are the physical symmetries associated with this entangling sphere $S$. We will refer to them as \emph{corner symmetries} and we call the associated algebra the \emph{corner symmetry algebra}. This nomenclature is adopted in order to distinguish them from \emph{boundary symmetries} with charges living on the whole time development of a time-like boundary like on Figure \ref{Delta} (or a null boundary), which we will come back to in a future publication. Differently from the boundary symmetry algebra, the corner symmetry algebra is \emph{independent} of the choice of boundary conditions. Moreover, the corner symmetry algebra is in a sense a subalgebra\footnote{More precisely, for a time-like boundary $\Delta$ we can associate a symmetry algebra $\mathfrak{g}_\Delta(S)$ to any sphere $ S \in \Delta$. If the sphere is in $\pa \Delta$ we recover the corner symmetry algebra. If $S$ is in the bulk of $\Delta$ the boost symmetry is broken down to an abelian subgroup of $\sll(2,\mathbb{R})_\per$ while the rest of the corner symmetry is still part of $\mathfrak{g}_\Delta(S)$.}    of the  boundary symmetry algebra. In that sense it is a \emph{universal} component of any boundary symmetry algebra and a fundamental component of any quantization of gravity. It is for these reasons that we focus our attention on it.

\begin{figure}[h!]
\centering
\begin{subfigure}[t]{74mm}
\centering
\includegraphics[height=39mm]{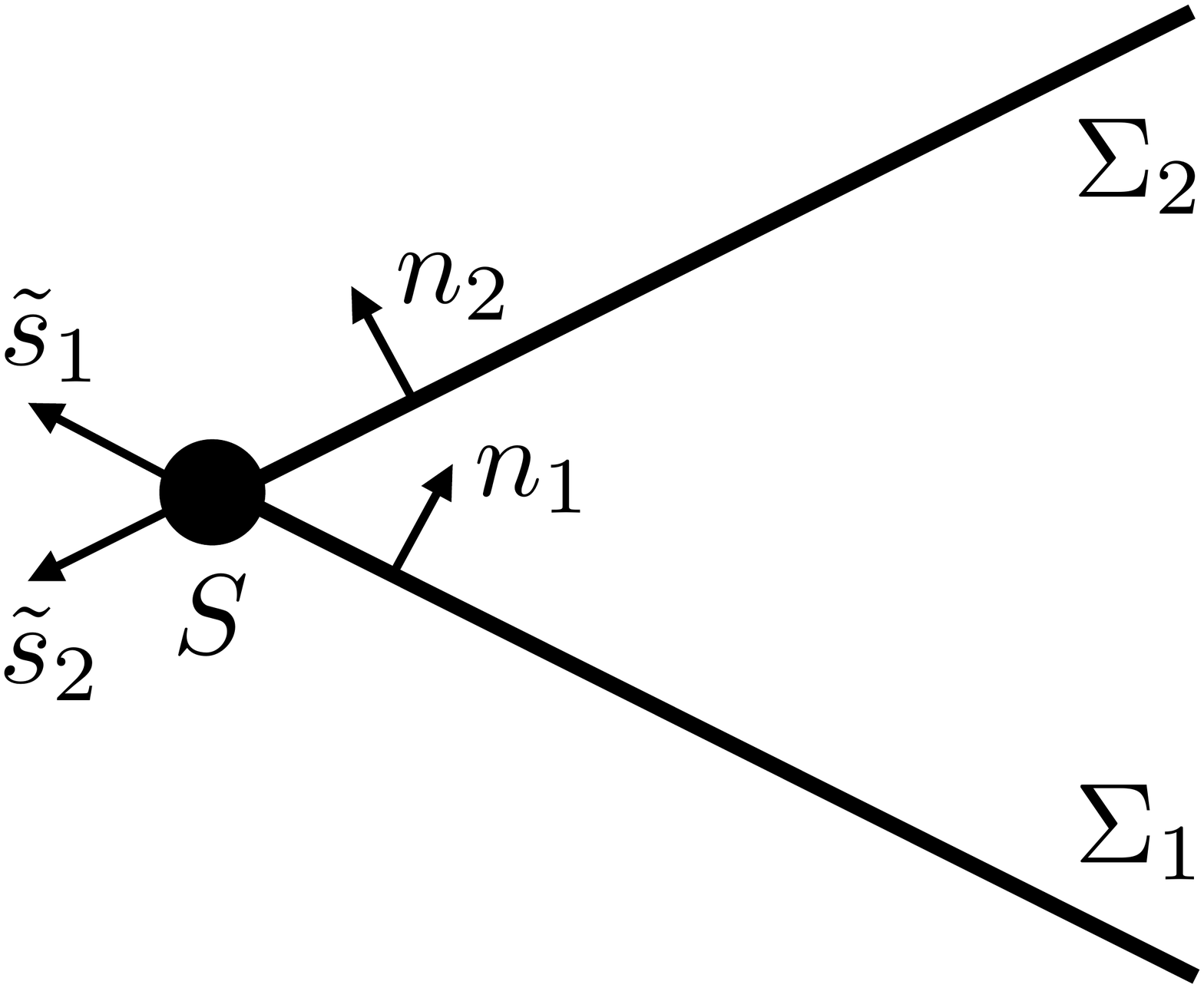}
\caption{Entangling wedge foliated by space-like Cauchy hypersurfaces $\Sigma$ all joining at the corner 2-sphere $S$.}\la{Wedge}
\end{subfigure}
\hspace*{14.mm}
\begin{subfigure}[t]{74mm}
\centering
\includegraphics[height=39mm]{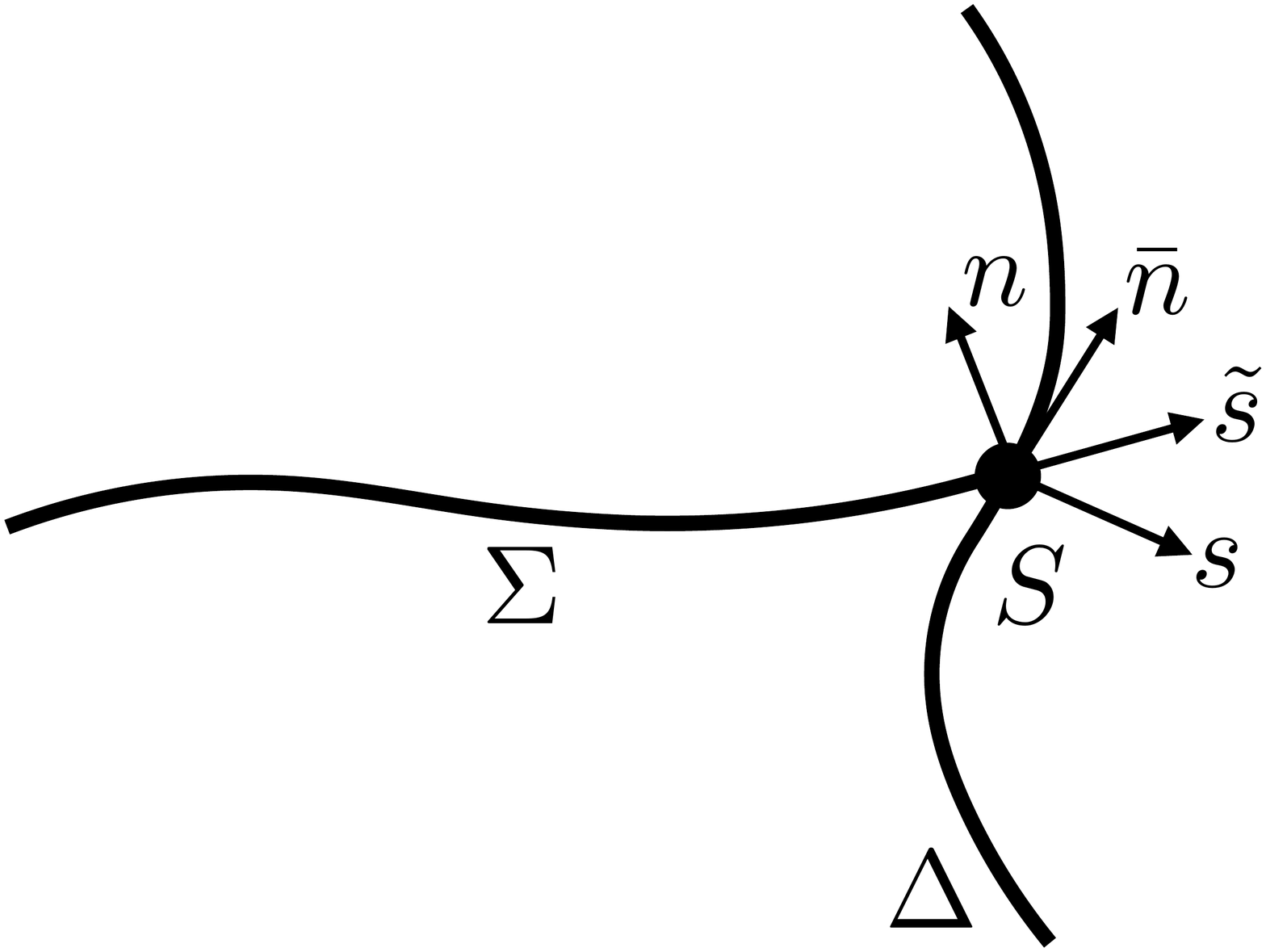}
\caption{Time-like boundary $\Delta$ intersecting a space-like Cauchy surface $\Sigma$ at a 2-sphere $S$.}\la{Delta}
\end{subfigure}
\caption{Cauchy surfaces $\Sigma$ with boundary sphere $S$ and associated sets of normals.}
\end{figure}\la{figure}

The covariant phase space formalism is notoriously plagued by so-called corner ambiguities \cite{Jacobson:1993vj,Iyer:1994ys,Julia:2002df}, which as the name suggests can potentially affect the corner charges and their symmetry algebra. We propose to view these not as ambiguities but rather as \textit{features}. 
This is achieved with a new prescription which assigns a unique pre-symplectic potential to a given Lagrangian. 
The fact that it is possible to have a non-ambiguous definition of the symplectic potential is due to the properties of the bigraded differential calculus that we have outlined in the Appendix \ref{appendix1}. 
From the resolution of the symplectic ambiguity, we derive the main message of the present work, which will then serve as a guide for the rest of this series of articles: Different formulations of gravity, which are equivalent in the bulk, generically differ by the presence of a corner term in their symplectic potential, and as such they have different corner symmetry algebras.
Once we know that there is a unique presymplectic  potential, 
we can  assign a  proper ``bulk-boundary decomposition'' of the symplectic potentials that mirror the bulk-boundary decomposition of Lagrangians. This explains why different formulations of the same theory, namely general relativity with diffeomorphism gauge symmetry, which have the same bulk symplectic structure, can have different corner symmetry algebras. Proceeding with the systematic study of the corner symplectic potentials gives us an organizing principle for understanding the corner symmetries. 
Moreover, this systematic treatment requires to acknowledge that boundary Lagrangians also posses their own symplectic potentials, which naturally live at corners\footnote{Let us clarify that, while the corner surface $S$ can be understood as a boundary of a boundary from the spacetime perspective, the pre-symplectic potential is defined on a given Cauchy slice $\Sigma$ only. In this case  $S$ has to be understood as simply the boundary of $\Sigma$. }
 \cite{Lee:1990nz,Harlow:2019yfa,Geiller:2019bti}.

In this paper we study general relativity in metric and tetrad variables. In the metric case, we consider the Einstein--Hilbert Lagrangian together with the canonical Lagrangian in ADM-like variables. Even though they describe the same bulk theory, these two Lagrangians lead to symplectic potentials which differ by a corner term (which we call the \textit{relative} potential), and to different representations of the corner symmetries. In particular, while both formulations have $\text{diff}(S)$ as a part of the corner symmetries, the Einstein--Hilbert formulation has an extra $\sll(2,\mathbb{R})$ corner symmetry \cite{Donnelly:2016auv}, called the boost symmetry and denoted $\sll(2,\mathbb{R})_\perp$, which is trivially represented in the canonical case. This takes us back to the question raised above: Which formulation of gravity has the maximal corner symmetry algebra? This question is important for quantum gravity, as quantizing this symmetry algebra provides important information about the Hilbert space. This idea is at the heart of loop quantum gravity (LQG) \cite{Ashtekar:2004eh,Perez:2012wv}. There, one considers the tetrad formulation of gravity represented by the Einstein--Cartan Lagrangian with the so-called topological Holst term \cite{Holst:1995pc}. This formulation has the advantage of having non-vanishing $\SU(2)$ charges associated with internal gauge transformations, which are nothing but the geometrical fluxes on which the whole LQG quantization in terms of spin network states rests. Applying our systematic study of the symplectic potential to the tetrad formulation of gravity reveals the same nesting structure: The tetrad formulations (i.e. with or without the Holst term, and with or without the imposition of the time gauge) differ from the canonical or Einstein--Hilbert metric formulations by a corner term in the symplectic potential \cite{DePaoli:2018erh}, and the knowledge of this corner term is crucial to the investigation of the corner symmetry algebra and the quantization of the theory. The detailed structure of the corner term of tetrad gravity will be investigated in the companion paper \cite{Edge-Mode-II}, and its application to the quantization in \cite{Edge-Mode-III,Edge-Mode-IV,DFMS}.

In the present work and in the companion paper \cite{Edge-Mode-II}, we therefore provide a systematic analysis of the symplectic potentials of various formulations of general relativity (which are all equivalent in the bulk), and explain how they correspond in turn to various ways in which the \textit{gauge} symmetries are turned into \textit{physical} symmetries at the corner. Table \ref{table} below summarizes these results, and displays the corner symmetries for the various formulations of gravity which we consider.
\begin{table}[h]
\centering
\begin{tabular}{|c|c|c|c|c|c|}
\hline
&\multicolumn{5}{|c|}{\textbf{Corner symmetries} $\mathfrak{g}^S$}\\
\hline
\textbf{Formulation of gravity}& $\mathrm{diff}(S)$ & $\sll(2,\mathbb{R})_\per$ & $\sll(2,\mathbb{R})_\para$ & $\;\mathfrak{su}(2)\;$ & $\mathrm{boosts}$ \\ 
\hline
Canonical general relativity (GR) & $\checkmark$ & & & & \\ 
\hline
Einstein--Hilbert (EH) & \checkmark &\checkmark & & & \\
\hline
Einstein--Cartan (EC) & \checkmark & & & & \checkmark \\
\hline
Einstein--Cartan--Holst (ECH) & \checkmark & &\checkmark & \checkmark & \checkmark \\
\hline
Einstein--Cartan--Holst + time gauge (ECHt) & \checkmark & &\checkmark & \checkmark & \\
\hline
\end{tabular}
\caption{Checkmarks denote which sectors of the corner symmetry algebra present a non-trivial representation in the given formulation of gravity. The symbols $\perp$ and $\parallel$ denote the fact that the corresponding $\sll(2,\RR)$ algebra is associated respectively to the normal and tangent parts of the metric at the corner. $\su(2)$ and boost denote the decomposition of the corner Lorentz symmetry algebra. For instance $\su(2)$ is trivially represented in Einstein--Cartan and non-trivially represented in Einstein--Cartan with the Holst term. }
\label{table}
\end{table}

Following this systematic investigation of the corner symplectic potentials, which leads to the results of Table \ref{table}, clearly shows that some formulations of gravity have more corner structure than others, and therefore a richer representation structure. The canonical formulation, which we label GR, provides the minimal bulk symplectic potential common to all the formulations. Any other formulation in Table \ref{table} has a symplectic potential which is the sum of the bulk potential $\theta_\GR$ and a corner term. This latter therefore clearly controls the part of the symmetry algebra which any formulation of gravity may have in addition to $\text{diff}(S)$.

Having established that different formulations of gravity have different corner symmetry algebras, and that this latter is controlled by the corner symplectic potential, we can then apply the edge mode formalism introduced in \cite{Donnelly:2016auv} and pushed further in \cite{Speranza:2017gxd,Geiller:2017xad,Geiller:2017whh,Geiller:2019bti,Setare:2018mii}. This consists in restoring the gauge-invariance broken by the presence of a boundary by adding edge mode fields. By doing so, the charges of physical symmetries become charges associated with transformations of the edge mode fields. While at the classical level this may seem like a simple reshuffling of information, the so-introduced edge modes cannot be dispensed with in the quantum theory, as we will show in details in the companion papers \cite{Edge-Mode-II, Edge-Mode-III, Edge-Mode-IV}.

We therefore have a clear roadmap for the study of edge modes in gravity. The first step, initiated in this paper, consists in carefully studying the decomposition of the symplectic potentials, and analyzing how the corner terms lead to inequivalent representations of the corner symmetry algebras. Then, edge modes can be introduced as a convenient parametrization of this corner symmetry algebra, which in addition has the advantage of restoring gauge-invariance. With these edge modes at hand, one can then address the issue of gluing of local subregions \cite{Donnelly:2016auv}, the reconstruction of holographic dynamics in terms of conservation laws for the symmetry charges \cite{Freidel:2016bxd,Freidel:2019ees,Freidel:2019ofr}, and most importantly the issue of quantization of the corner degrees of freedom \cite{Freidel:2015gpa,Freidel:2019ees}.

This paper is organized as follows. Section \ref{sec:metric} is devoted to general relativity in metric variables. There, we study three Lagrangians and analyse in details the relationship between their symplectic potentials. Using the covariant phase space formalism we then investigate the corner symmetry algebra for diffeomorphisms tangent to the corner sphere $S$. We show that the corner symmetry for the ADM formulation of gravity is simply $\text{diff}(S)$, and recall \cite{Donnelly:2016auv} that the Einstein--Hilbert corner symmetry algebra is\footnote{$G^S$ denotes the sets of maps $S\to G$.} 
$\text{diff}(S)\ltimes\sll(2,\mathbb{R})_\perp^S$, where the $\sll(2,\mathbb{R})_\perp^S$ stands for the local boost transformations of $S$. We also prove that the difference in the corner symmetries stems from the corner potential. We conclude this section by showing how completely analog results hold for the Gibbons--Hawking Lagrangian when considering a time-like boundary.

Having established these results, we then move in Section \ref{sec:tetrad} to the study of tetrad gravity. There, we first focus on BF theory, and then introduce Einstein--Cartan (--Holst) gravity (ECH) as a constrained BF theory. As it turns out, the superficial analysis of the corner symmetries of tetrad gravity, which we recall in Section \ref{subsec:ECH}, reveals the algebra $\text{diff}(S)\ltimes\sll(2,\mathbb{C})^S$, where the $\sll(2,\mathbb{C})$ is due to internal Lorentz symmetries, and the boost component $\sll(2,\mathbb{R})_\perp$ is absent (this last observation was first noted in \cite{Jacobson:2015uqa} and further analyzed in \cite{Prabhu:2015vua,DePaoli:2018erh,Oliveri:2019gvm,aneesh2020law}). Compared with the metric case, much more work is required in order to decompose the potential of Einstein--Cartan--Holst gravity in terms of the fundamental bulk piece $\theta_\GR$ and a corner potential. This study is therefore deferred to \cite{Edge-Mode-II}, where we show that a careful analysis of the corner symplectic structure reveals the presence of an additional $\sll(2,\mathbb{R})_\para$ symmetry algebra, distinct from the boost symmetry algebra which was denoted $\sll(2,\mathbb{R})_\perp$ above. This algebra is in fact that of the tangential metric on $S$, and its quantization reveals that the area spectrum is discrete. This illustrates the kind of important information encoded in the corner symplectic potential, and why we devote \cite{Edge-Mode-II} to its detailed analysis.

In order to make the paper as self-contained as possible, we have included appendices containing many technical details and full derivations. Appendix \ref{appendix1} contains a presentation of the covariant phase space formalism. The subsequent appendices gather various proofs and details of calculations used throughout the main text.

\section{Metric gravity}
\label{sec:metric}

Let us start with the metric formulation of gravity. Our goal is to gather familiar results and to reinterpret them in light of the organizing principle mentioned in the introduction: Different formulations of gravity have different symplectic potentials, which differ only by corner terms, and as such lead to different corner charges and symmetry algebras. This is the content of the first two lines of Table \ref{table}. Metric gravity is the simplest and most familiar setup in which one can appreciate this result. This will also give us  the opportunity to introduce some notations and conventions, and to set the stage for the analysis of tetrad gravity initiated here and continued in the companion paper \cite{Edge-Mode-II}.

Throughout this paper we consider a spacetime $M$ equipped with a Lorentzian metric $g_{\mu\nu}$ of signature $(-,+,+,+)$, and denote its volume form by\footnote{We chose the orientations $\rd^4x=\rd x\wedge\rd y\wedge\rd z\wedge\rd t$ and $\rd^3x=\rd x\wedge\rd y\wedge\rd z$. For simplicity we will drop all the integration measures such as $\rd^4x$ or $\rd^3x$ when writing integrals below.} $\eps\coloneqq\sqrt{|g|}\,\rd^4x$. The $3+1$ decomposition involves the choice of a foliation, with codimension-1 space-like slices $\Sigma$ such that $M=\Sigma\times\mathbb{R}$. Each slice has a normal 1-form $\un=n_\mu\rd x^\mu$ which is, up to a scale, intrinsically attached to the hypersurface. For a space-like surface we normalize $\un$ by demanding that it satisfies $g^{\mu\nu}n_\mu n_\nu=-1$. We denote the normal vector, which is metric-dependent, by $\hat{n}=n^\mu\partial_\mu$, and choose it to be outward-pointing. This distinction is important since $\hat{n}$ and $\un$ behave differently under variations. Some of the Lagrangians considered here depend on the pair, which we simply denote $n \coloneqq(\un,\hat{n})$. The slice has an induced metric $\tilde{g}_{\mu\nu}\coloneqq g_{\mu\nu}+n_\mu n_\nu$. The volume form on $\Sigma$, denoted by $\teps$, is related to the spacetime volume form by $\eps=\un\wedge\teps$.
For a time-like normal this means that\footnote{We use the notation $(\hat v\ip\omega)_{b_1\cdots b_{p-1}}\coloneqq v^a\omega_{ab_1\cdots b_{p-1}}$ for any $p$-form $\omega_{b_1\cdots b_{p}}$.} $\teps=-\hat{n}\ip\eps$ and $\teps=\pm\sqrt{|\tilde{g}|}\,\rd^3x$, where the sign depends on 
whether $\Sigma$ is a future $(+)$ or past $(-)$ boundary of $M$. Finally, we will also use the notations $\eps_\mu=\partial_\mu\ip\eps$ for a basis of  codimension-1 forms, and $\teps_\mu=\partial_\mu\ip\teps$ for a basis of codimension-2 forms with $n^\mu\teps_\mu=0$. These enter the Stokes theorem as
\be
\int_M \epsilon\nabla_\mu v^\mu=\int_M\rd(v^\mu\epsilon_\mu)=\int_{\pa M}\epsilon_\mu v^\mu=\int_{\pa M}\teps\,n_\mu v^\mu,
\ee
and similarly for integration over $\Sigma$ with $(\eps,\eps_\mu)$ replaced by $(\teps,\teps_\mu)$.

\subsection{Metric Lagrangians}

It is well known that  given a Lagrangian $L$,
one determines  unambiguously from it its equations of motion $E_L$.
It is also repeatedly emphasized \cite{Lee:1990nz,Iyer:1994ys,Jacobson:1993vj, Wald:1999wa}, that the symplectic potential $\theta_L$, which is a co-dimension one form linear in the field variation and  satisfies 
\be \label{var}
\delta L =E_L +\rd \theta_L,
\ee
is  ambiguously defined. It can be shifted by corner terms $ \theta_L \to \theta_L +\rd \vartheta$ where $\vartheta$ is a co-dimension 2 form linear in the field variations.
In this work we propose a  new way to resolve this ambiguity.
It turns out that this ambiguity comes about if one decides to use \eqref{var} as a \emph{definition} of the symplectic potential.
We take a different route,   and, given $L$, we use the general bi-graded differential calculus to give an explicit construction of both $E_L$ and $\theta_L$ such that \eqref{var} follows from the definition. 
The detailed formula of $\theta_L$  for a general Lagrangian $L$ is given in Appendix  \ref{appendix1}.
This proposal means that we have an unambiguous definition of $\theta_L$  given a Lagrangian $L$ \emph{and} a choice of field coordinates. 
It also means that different Lagrangians that possess the same equations of motions can still be distinguished by their symplectic potential. 
The fact that such a procedure is possible is not surprising when the Lagrangians are first order. For instance if  $L= \eta_{AB}\rd \phi^A \wedge  *\phi^B + V(\phi^A)$ for an arbitrary potential $V$ continuing no derivatives, our general proposal simply replaces derivatives by variations and assigns   $\theta_L = \eta_{AB}\delta \phi^A \wedge  *\phi^B$.
The addition of a corner term to the symplectic potential is encoded in the addition of a boundary Lagrangian.
It is not viewed as an ambiguity. 
Lets assume that $L_1$ and $L_2$ are two Lagrangians that share the same equations of motion. This means that they differ by a boundary lagrangian $L_2= L_1+ \rd \ell_{2/1}$.
According to our general prescription we then have  two symplectic potentials $\theta_1:=\theta_{L_1}$ and $\theta_2:=\theta_{L_2}$. By construction they differ by a corner term $ \theta_2 = \theta_1 + \rd \vartheta_{2/1}$.
The corner term\footnote{We also establish in Appendix \ref{Noethera} that the correspondence $\ell_{1/2} \to \vartheta_{1/2}$ is uniquely defined and that \eqref{correspondence} is a consequence of our definition, not a definition.} arises from the variation of the boundary Lagrangian.  
\be\label{correspondence}
\delta \ell_{2/1} = \theta_2-\theta_1 + \rd \vartheta_{2/1}. 
\ee
In other words, the corner term is simply the symplectic potential of the boundary Lagrangian.
The general formulae are given in Appendix \ref{Noethera} and will be expanded on in \cite{GeneralizedL}. Here we illustrate this philosophy on a set of gravitational examples. We introduce three different gravitational Lagrangians, the Einstein-Hilbert (EH) Lagrangian,
the canonical gravity Lagrangian (GR) and the Gibbons-Hawking (GH) Lagrangian and study their relationship, establishing in each case the general formula \eqref{correspondence}.
Note that the formula was also established in \cite{Harlow:2019yfa}
for  the relation between the EH  and GH Lagrangians but presented there as a relations between boundary conditions instead.

Let us start our discussion by grounding it in the choice of a Lagrangian describing the theory. As is well-known, there exist many alternative Lagrangians for metric gravity\footnote{By this we mean Lagrangians which lead to Einstein's equations of general relativity for a spin 2 field, and not any modified theory of gravity.}. The most popular  choice is  the Einstein--Hilbert Lagrangian. In units where $8\pi G=1$, it is given by
\be\label{EH Lagrangian}
L_\EH[g]=\f{1}{2}\eps R.
\ee
The variation of this Lagrangian reads
\be
\delta L_\EH[g]=\f{1}{2}\eps\left(R_{\mu\nu}-\f{1}{2}Rg_{\mu\nu}\right)\delta g^{\mu\nu}+\eps\nabla_\mu\theta^\mu_\EH,
\ee
where the first term identifies the equations of motion, and the second term depends on the pre-symplectic potential.
The symplectic potential uniquely associated with the Einstein-Hilbert according to the general procedure design in appendix \ref{Noethera} is the usual expression
\be\label{EH current}
\theta^\mu_\EH\coloneqq\f{1}{2}(g^{\alpha\beta}\delta\Gamma^\mu_{\alpha\beta}-g^{\alpha\mu}\delta\Gamma^\beta_{\alpha\beta}).
\ee
This latter serves as the starting point for the construction of the covariant phase space formalism, which we recall in Appendix \ref{appendix1}, and the derivation of the corner charges and symmetry algebra. We are going to focus on this metric potential, show that it contains a corner term, and draw from this simple fact important conclusions about the corner symmetries of gravity.

Having obtained the pre-symplectic potential \eqref{EH current} from the Lagrangian \eqref{EH Lagrangian}, it is natural to ask, already at this point, what happens if one starts from a different Lagrangian. Since we have at our disposal the vector field $\hat{n}$, one can use the Gauss--Codazzi equation relating the 4-dimensional Ricci scalar $R$ to the Ricci scalar $\tilde{R}$ of the slice $\Sigma$ and its extrinsic curvature tensor $\tK_{\mu\nu}={\tilde{g}_\mu}^\alpha{\tilde{g}_\nu}^\beta\nabla_\alpha n_\beta$. It enables us to rewrite the Einstein--Hilbert Lagrangian \eqref{EH Lagrangian} in the form \cite{Wald:1984rg,Hawking:1996ww,blaunotes}
\be\label{EH=GR+K}
L_\EH[g]=L_\GR[\tilde{g}, {n}]+{\rd} L_{\EH/\GR}[\tilde{g}, {n}],
\ee
where
$L_{\GR}$ is a bulk Lagrangian and $L_{\EH/\GR}$ is a boundary Lagrangian given by
\be\label{GR Lagrangian}
L_\GR[\tilde{g}, {n}]\coloneqq\f{1}{2}\eps\Big(\tilde{R}-(\tilde{K}^2-\tilde{K}^{\mu\nu}\tilde{K}_{\mu\nu})\Big),\q L_{\EH/\GR}[\tilde{g}, {n}]\coloneqq\eps_\mu\big(n^\mu\tilde{K}-\tilde{a}^\mu\big).
\ee
We call the  boundary Lagrangian the ``relative Lagrangian'' between the two formulations EH and GR. It is a codimension-1 form built with the trace $\tilde{K}=\nabla_\mu n^\mu$ of the extrinsic curvature tensor and the acceleration vector $\tilde{a}^\mu=n^\alpha\nabla_\alpha n^\mu$. The Lagrangian $L_\GR$ is sometimes referred to as the ADM Lagrangian. For reasons which will become clear below when studying the symplectic potentials, we have chosen to simply call it GR, for general relativity. Using the variational identity \eqref{variation induced K}, one finds that the pre-symplectic potential derived from the Lagrangian $L_\GR$ is
\be
\theta^\mu_\GR=-\f{1}{2}n^\mu(\tK\tilde{g}^{\alpha\beta}-\tK^{\alpha\beta})\delta\tilde{g}_{\alpha\beta}+\tilde{\theta}^\mu_\GR,
\ee
where the last piece is such that $n_\mu\tilde{\theta}^\mu_\GR=0$ and therefore irrelevant when integrated on the slice $\Sigma$. This potential encodes the symplectic structure encountered in the canonical (or Hamiltonian) decomposition of general relativity \cite{Arnowitt:1959ah,Arnowitt:1962hi}, and expresses the well-known fact that
\be 
\tilde{P}^{\mu\nu}\coloneqq\teps(\tK\tilde{g}^{\mu\nu}-\tK^{\mu\nu})
\ee
is the momentum density conjugated to the induced metric $\tilde{g}_{\mu\nu}$. In the absence of matter, this momentum satisfies the conservation equation
\be
\tilde{\nabla}_\mu\tilde{P}^{\mu\nu} =0,
\ee
where $\tilde{\nabla}$ is the induced Levi--Civita connection on $\Sigma$. This conservation equation is nothing but the spatial diffeomorphism constraint.

By distinguishing the EH and GR Lagrangians, which differ by only a corner term and therefore reproduce the same bulk equations of motion of general relativity, we have obtained two different pre-symplectic potentials. The rest of this section is devoted to analyzing in depth the relationship between them. This will explain in particular, as stated in Table \ref{table}, how they lead to two inequivalent representations of the corner symmetries of gravity.

Let us conclude this part with an important observation. Above we have used the time-like normal $n$ in order to decompose the metric $g$ of the EH formulation into the variables $(\tilde{g},n)$ leading to the GR Lagrangian. However, we could have equally well considered a space-like normal $s=(\us,\hat{s})$, and the associated induced metric $\bar{g}_{\mu\nu}\coloneqq g_{\mu\nu}-s_\mu s_\nu$ and geometrical quantities to decompose
\be\label{EH=GH+K}
L_\EH[g]=L_\GH[\bar{g},s]+\rd L_{\EH/\GH}[\bar{g},s],
\ee
where now the bulk and boundary Lagrangians are
\be\label{GH Lagrangian}
L_\GH[\bar{g},s]\coloneqq\f{1}{2}\eps\Big(\bar{R}+(\bar{K}^2-\bar{K}^{\mu\nu}\bar{K}_{\mu\nu})\Big),\q L_{\EH/\GH}[\bar{g},s]\coloneqq-\eps_\mu\big(s^\mu\bar{K}-\bar{a}^\mu\big).
\ee
$L_\GH$ is a third possible bulk Lagrangian which one can consider for general relativity. We have named it GH for Gibbons--Hawking, since when we pull back $-L_{\EH/\GH}$ on a on a time-like boundary $\Delta$ with normal $\hat{s}$ (see Figure \ref{Delta}), the acceleration term vanishes since $\bar{a}^\mu s_\mu=0$ and we are left with the Gibbons--Hawking term \cite{Gibbons:1976ue} $\int_\Delta\bar{\eps}\bar{K}$.

The three Lagrangians $L_\EH$, $L_\GR$, and $L_\GH$, are natural starting points for our study of the covariant phase space and corner symmetries. 
Lets us first focus on the comparison between the Einstein--Hilbert $L_\EH$ and canonical $L_\GR$ Lagrangians, and consider only space-like foliations with normal $\hat{n}$. For the time being, we are therefore left with the task of understanding the physical meaning of the difference between the two pre-symplectic potentials $\theta_\EH$ and $\theta_\GR$ derived from the two Lagrangians $L_\EH$ and $L_\GR$. Once we understand this difference, which is related to the presence of a corner term, we will be able to move on to tetrad gravity, and compare these two metric potentials to the potential of tetrad gravity. This way, we are building a systematic study of the potentials of various formulations of gravity.

\subsection{Symplectic potentials}
\la{sec:2.1.1}

The pre-symplectic potentials are obtained by integrating the currents along the slice $\Sigma$. After a slight rewriting, the pre-symplectic potential for EH gravity is found to be
\be\label{EHpotential}
\Theta_\EH=\int_\Sigma\tilde{\eps}\,n_\mu\theta^\mu_\EH=\f{1}{2}\int_\Sigma\tilde{\eps}\,n^\mu\nabla^\nu(\delta g_{\mu\nu}-g_{\mu\nu}g^{\alpha\beta}\delta g_{\alpha\beta}),
\ee
and that coming from the GR Lagrangian is
\be
\Theta_{\GR}=\int_{\Sigma}\tilde{\eps}\,n_\mu\theta^\mu_{\GR}=\f{1}{2}\int_{\Sigma}\tilde{\eps}(\tK\tilde{g}^{\mu\nu}-\tK^{\mu\nu})\delta\tilde{g}_{\mu\nu}.
\ee
We therefore have at our disposals two natural pre-symplectic potentials for the same bulk theory, which is metric gravity.

The reason for which we have decided to label the second symplectic potential by GR, for general relativity, is that one can think of it as the ``fundamental''   potential capturing the bulk canonical degrees of freedom which are common to any formulation of gravity. It is the canonical symplectic potential commonly used in the Hamiltonian analysis of gravity. Any other formulation of gravity can be understood as being built from this bulk GR potential plus some specific corner term.

Let us start by establishing this result in the case of the metric formulation of gravity. There, the statement is simply that the two potentials $\Theta_\EH$ and $\Theta_\GR$ introduced above differ only by a corner symplectic potential and a total field-space variation. A proof is given in \cite{Freidel:2013jfa, Brown:2000dz,Harlow:2019yfa} and recalled in Appendix \ref{appendix:EH potential} for completeness. Explicitly, we have that
\be\label{EH-GR potentials relation}
\Theta_\EH=\Theta_\GR+\Theta_{\EH/\GR}-\delta\left(\int_\Sigma\tilde{\eps}\tK\right),\qquad 
\Theta_{\EH/\GR}=\int_S\theta_{\EH/\GR}.
\ee
The last term in the first equality implements a canonical transformation, and its presence does not affect the symplectic form $\Omega_\EH=\delta\Theta_\EH$ since $\delta^2=0$. The second term on the right-hand side is the corner symplectic potential of the Einstein--Hilbert formulation of gravity, it is expressed as the integral of the relative symplectic potential along the corner  $S$. As shown in Appendix \ref{K-appendix} (see equation \eqref{Cmain}), the relation \eqref{EH-GR potentials relation} can also be expressed in terms of the boundary Lagrangian \eqref{GR Lagrangian} as
\be\label{Cmain}
\delta L_{\EH/\GR}+ {\rd}\theta_{\EH/\GR}=\theta_\EH-\theta_\GR,
\ee
showing how the boundary Lagrangian affects the form of the  corner symplectic potential, in agreement with \cite{Lee:1990nz,Harlow:2019yfa,Geiller:2019bti}. As expected, this variation is naturally of the form $\delta L=E_L+ {\rd}\theta_L$, with $E_L$ the equations of motion and $\theta_L$ the corner potential. Because of the conventions which we have chosen when defining the relative potential, we have $\theta_{L_{\EH/\GR}}=-\theta_{\EH/\GR}$.

The corner symplectic form $\Omega_{\EH/\GR}=\delta\Theta_{\EH/\GR}$ derived from the corner potential does not vanish. Since the symplectic form encodes the phase space variables, it means that gravity in the EH  formulation differs from gravity in the canonical GR formulation by the presence of additional corner degrees of freedom. As can be seen in \eqref{mainEH-GR}, the explicit expression for the corner potential is\footnote{We give here once and for all these various equivalent formulae. Below and in \cite{Edge-Mode-II} we use whatever is more convenient depending on the calculation being performed.}
\be\label{corner EH potential}
\Theta_{\EH/\GR}=\int_S\teps_\mu\delta n^\mu_\perp=-\int_S\bar{\tilde{\eps}}\,\tilde{s}_\mu\delta n^\mu_\perp=\f{1}{2}\int_S\sqrt{q}\,\tilde{s}_\mu\delta n^\mu=-\f{1}{2}\int_S\eps^\per_{\mu\nu}n^\mu\delta n^\nu,
\ee
where
\be
\delta n^\mu_\perp\coloneqq\f{1}{2}(\delta n^\mu+g^{\mu\nu}\delta n_\nu),\q\bar{\tilde{\eps}}=-\hat{\tilde{s}}\ip\hat{n}\ip\eps,\q\eps^\per_{\mu\nu}\coloneqq\sqrt{q}(n_\mu\tilde{s}_\nu-\tilde{s}_\mu n_\nu),
\ee
with $\tilde{s}^\mu$ a space-like vector normal to $S$ and $n_\mu$ (see Figure \ref{Wedge}), and $|\bar{\tilde{\eps}}|=\sqrt{q}$ with $q_{\mu\nu}=\tilde{g}_{\mu\nu}-\tilde{s}_\mu\tilde{s}_\nu$ the induced metric on $S$. Again, the derivation of this result is recalled in Appendix \ref{appendix:EH potential}. Taking into account the presence of this corner symplectic potential is of crucial importance in order to properly describe the boundary gravitational degrees of freedom. It tells us that the EH formulation has the additional canonical pair $(\sqrt{q}\,\tilde{s}_\mu,n^\mu)$ living at the corner.

\subsection{Gibbons--Hawking Lagrangian and relative boost}

The analysis done for a space-like surface $\Sigma$ with normal form $\un$ can be reproduced effortlessly for a time-like surface $\Delta$ with normal form $\underline{s}$. Similarly to \eqref{EH-GR potentials relation}, one finds that
\be
\int_\Delta\theta_\EH=\int_\Delta\theta_\GH+\int_{\pa\Delta}\theta_{\EH/\GH}-\delta\left(\int_\Delta \bar{\eps}\bar{K}\right),
\ee
where 
\be
\theta_{\GH} = \bar{P}^{\mu\nu} \delta \bar{g}_{\mu\nu}, \qquad
\theta_{\EH/\GH} = \bar{\eps}_\mu\delta s^\mu_\perp,
\ee
and where the bulk and boundary canonical variables are
\be 
\bar{P}^{\mu\nu}\coloneqq\bar{\eps}(\bar{K}^{\mu\nu}-\bar{K}\bar{g}^{\mu\nu}),
\qquad
\delta s^\mu_\perp \coloneqq \f{1}{2}(\delta s^\mu+g^{\mu\nu}\delta s_\nu), \qquad
\bar{\eps}_\mu = \pa_\mu\ip\hat{s}\ip \epsilon.
\ee
We see that the bulk-boundary decomposition in the case of a time-like boundary is exactly the same as in the space-like case, with the simple replacement $(\Sigma,{n},\tilde{g},\tilde{K})\to(\Delta, {s}, \bar{g},\bar{K})$. The fact that the mathematical structures of the time-like and space-like cases are similar suggests that we can draw an analogy between the two cases, even if their conceptual interpretation is quite different. For instance, the time-like analog of the diffeomorphism constraint is the energy-momentum conservation
\be
\nabla_\mu\bar{P}^{\mu\nu}=0, 
\ee
which suggests that the time-like boundary has the structure of an hydrodynamical fluid \cite{Thorne:1986iy, Bhattacharyya:2008jc,Compere:2012mt}. The time-like analog of the notion of a state, which is, according to our semi-classical correspondence, a Lagrangian subspace of the bulk symplectic structure, is simply a choice of boundary condition, which can also be defined as a Lagrangian subspace of the ``time-like symplectic structure'' $\int_{\Delta}\delta\theta_{\GH}$. These analogies play of course a key role in the formulation of the AdS/CFT correspondence for time-like boundaries near infinity of asymptotic AdS \cite{deBoer:1999tgo, Marolf:2004fy, Skenderis:2008dg, Lawrence:2006ze}. The essential point is that some of the key notions used there survive at finite distance. For instance, in 3d the full quantum gravity solution can be reconstructed at finite distance by pushing this analogy to its limits and understanding the radial motion as a $T\bar{T} $ deformation \cite{Freidel:2008sh,Mazenc:2019cfg,Tolley:2019nmm, Iliesiu:2020zld}.

Now that we understand the two corner  potentials $\Theta_{\EH/\GR}$ and $\Theta_{\EH/\GH}$ we can evaluate their difference\footnote{Throughout this work and in \cite{Edge-Mode-II}, given two formulations A and B we always denote the relative potential by $\Theta_\text{A/B}=\Theta_\text{A}-\Theta_\text{B}$, and similarly for the relative Lagrangian and charge. This leads to useful transition formulae of the form $\Theta_\text{A/C}=\Theta_\text{A/B}-\Theta_\text{C/B}=\Theta_\text{B/C}-\Theta_\text{B/A}$.} and obtain the relative potential between the GR and GH formulations. We can view the corner $S$ as both a boundary of $\Sigma$  and  of $\Delta$ with the pairs of normals shown in Figure \ref{Delta}. In the GR formulation, the corner $S$ is characterized by the orthornormal pair  $(n_\mu,\tilde{s}_\mu)$ with $n_\mu$ time-like. Conversely, in the GH formulation the corner is characterized by $(s_\mu,\bar{n}_\mu)$ with ${s}_\mu$ space-like. As shown in Appendix \ref{K-appendix}, the relative potentials between these various formulations satisfy
\be\la{T-GH/GR}
\Theta_{\GH/\GR}= \Theta_{\EH/\GR}-\Theta_{\EH/\GH}=\int_S\big(\teps_\mu\delta n^\mu_\perp-\bar\eps_\mu\delta s^\mu_\perp\big)
\ee
To evaluate this we introduce the boost angle $\eta$ defined by $\hat{n}\cdot\hat{s}=\sinh\eta$. If  this angle is fixed on $S$, the GR and GH formulations lead to the same symplectic potential. If the boost angle is allowed to vary  we obtain (details in Appendix \ref{K-appendix}) that the relative potential has the Regge form 
\be
\Theta_{\GH/\GR}=\int_S\tilde{\bar{\eps}}\,\delta\eta.
\ee
This expression embodies the fact that the boost angle is conjugated to the area form at the corner, which was first established in the discrete context by Regge \cite{Regge} and in the continuum by Hayward \cite{Hayward:1993my}, and used to get insights into quantum black hole physics \cite{Carlip:1993sa,Massar:1999wg}. One can view this canonical pair as descending from the boost algebra $\sll(2,\mathbb{R})_\per$ after symmetry breaking induced by the presence of a time-like boundary \cite{Takayanagi:2019tvn}.

Now that we have established in these first examples that different formulations of gravity share the same bulk symplectic potential but differ at the corner, we can investigate the consequence of this fact for the representation of the corner symmetry algebras and degrees of freedom. In the rest of this section we show that the additional corner potential (or the associated degrees of freedom) which differentiates the EH and GR formulations has two effects: It leads to non-trivial corner charges for transformations known as surface boosts, and it allows to ``covariantize'' the canonical GR formulation of gravity, in a way which we will explain below.

\subsection{Hamiltonian charges}

In this section we explain how the corner symplectic potential which differentiates the EH and GR formulations of gravity leads to a difference in the Hamiltonian boundary charges associated with diffeomorphisms. As usual for the discussion of diffeomorphisms in the covariant phase space formalism, we focus on the case of diffeomorphisms $\xi$ tangent to $\Sigma$, i.e. such that $\xi^\mu n_\mu=0$.

Charges are constructed by contracting the field variation, given by the  Lie derivative along a vector field, with the symplectic form. In the case of canonial gravity the symplectic form is
\be
\Omega_\GR=\delta\Theta_\GR=\f{1}{2}\int_\Sigma\delta\tilde{P}^{\mu\nu} \wedge \delta\tilde{g}_{\mu\nu}.
\ee
As is well-known, for  diffeomorphisms tangent to $\Sigma$, only those which are also tangent to the boundary sphere $S$, i.e. generated by vector fields $\xi=\xi^\mu\partial_\mu$ such that $\xi^\mu\tilde{s}_\mu\stackrel{S}=0$, are integrable and have a canonical generator. We therefore restrict our analysis to this case. We get (see Appendix \ref{appendix:Brown-York}) that the contraction of a Lie derivative with the GR symplectic form is an exact variational form given by
\be\label{vector constraint generator}
-\CL_\xi\ipp\Omega_\GR=\f{1}{2}\int_\Sigma\left(\delta\tilde{P}^{\mu\nu} \CL_\xi\tilde{g}_{\mu\nu}-\CL_\xi\tilde{P}^{\mu\nu} \delta\tilde{g}_{\mu\nu}\right)=\delta\CH_\GR[\xi],
\ee
where the Hamiltonian generator is
\be
\CH_\GR[\xi]\coloneqq\f{1}{2}\int_\Sigma\CL_\xi\tilde{g}_{\mu\nu}\tilde{P}^{\mu\nu} .
\ee
As usual, this bulk integral can be integrated by parts to write
\be
\CH_\GR[\xi]=\CH_\GR^\Sigma[\xi]+\CH_\GR^S[\xi],
\ee
where we have introduced the bulk constraint $\CH_\GR^\Sigma[\xi]$ and the Hamiltonian charge $\CH_\GR^S[\xi]$. By construction of the covariant phase space, the bulk piece vanishes on-shell. It is nothing but the smeared vector constraint of canonical gravity, i.e.
\be
\CH_\GR^\Sigma[\xi]\coloneqq-\int_\Sigma\xi_\mu\tilde{\nabla}_\nu\tilde{P}^{\mu\nu} \approx0.
\ee
The charge is the piece which is left on-shell. It is given by a surface integral which here takes the form
\be\label{BY charge}
\CH_\GR^S[\xi]&\coloneqq\int_\Sigma\tilde{\nabla}_\mu(\tilde{P}^{\mu\nu} \xi_\nu)=\int_S\sqrt{q}\,\tilde{s}_\mu\xi_\nu(\tK\tilde{g}^{\mu\nu}-\tK^{\mu\nu})=-\int_S\sqrt{q}\,\tilde{s}_\mu\xi_\nu\tK^{\mu\nu}\cr
&=\int_S\sqrt{q}\,\tilde{s}_\mu n_\nu\nabla^\mu\xi^\nu\,,
\ee
where for the last two equalities we have used \eqref{induced derivative n} and the fact that $\xi$ is tangent to both $\Sigma$ and $S$. This surface integral is known as the Brown--York charge \cite{Brown:1992br}. It is important to notice for what follows that this charge does not depend on derivatives\footnote{Before integrating by parts.} of the vector field $\xi$. As we will explain shortly, this means that surface boosts \cite{Donnelly:2016auv} are represented trivially in the algebra of the Brown--York charges. The corner symmetry algebra associated with the GR action is therefore simply the diffeomorphism algebra $\mathfrak{g}_\GR^S=\text{diff}(S)$.

Let us now focus on the corner term $\Theta_{\EH/\GR}$ which differentiates the EH and GR symplectic potentials. This corner term gives a corner symplectic structure
\be
\Omega_{\EH/\GR}=\f{1}{2}\int_S\delta(\sqrt{q}\,\tilde{s}_\mu)\wedge \delta n^\mu.
\ee
The presence of this corner symplectic structure means that the boundary charges in the EH and GR formulations also differ. One is then lead to consider the difference between these two charges, which we denote by
\be \label{relative-charge}
\CH_{\EH/\GR}^S[\xi]\coloneqq\CH_\EH^S[\xi]-\CH_\GR^S[\xi].
\ee
This quantity, which we call the \textit{relative charge}, comes entirely from the new canonical pair $(\sqrt{q}\,\tilde{s}_\mu,n^\mu)$ living at the corner. In order to evaluate it, we contract the corner symplectic structure with a diffeomorphism. This gives
\be
-\CL_\xi\ip\Omega_{\EH/\GR}=\delta\CH_{\EH/\GR}^S[\xi],
\ee
where\footnote{
Similarly, the relative corner potential \eqref{T-GH/GR} between the $\GH$ and the $\GR$ formulations yields 
the relative charge  
\be\label{relative-charges2}
\CH_{\GH/\GR}^S[\xi]=\int_S\CL_\xi\eta\,\tilde{\bar{\eps}}.
\ee}
\be
\CH_{\EH/\GR}^S[\xi]=\f{1}{2}\int_S\sqrt{q}\,\tilde{s}_\mu\CL_\xi n^\mu.
\ee
It is very informative to rewrite this relative charge using the hypersurface orthogonality condition $\tilde{\nabla}^\mu n^\nu=\tilde{\nabla}^\nu n^\mu$. Indeed, this enables us to write
\be
\f{1}{2}\int_S\sqrt{q}\,\tilde{s}_\mu\CL_\xi n^\mu
&=\f{1}{2}\int_S\sqrt{q}\,\tilde{s}_\mu(\xi_\nu\nabla^\nu n^\mu-n_\nu\nabla^\nu\xi^\mu)\cr
&=\f{1}{2}\int_S\sqrt{q}\,\tilde{s}_\mu(\xi_\nu\nabla^\mu n^\nu-n_\nu\nabla^\nu\xi^\mu)\cr
&=-\f{1}{2}\int_S\sqrt{q}\,\tilde{s}_\mu n_\nu(\nabla^\mu\xi^\nu+\nabla^\nu\xi^\mu).
\ee 
We then see that the EH charge $\CH_\EH^S[\xi]=\CH_\GR^S[\xi]+\CH_{\EH/\GR}^S[\xi]$ is, as expected, the Komar charge
\be\la{Komar}
\CH_\EH^S[\xi]\coloneqq-\f{1}{2}\int_S\eps^\per_{\mu\nu}\nabla^\mu\xi^\nu.
\ee
This is indeed the result which we would have obtained by contracting the EH symplectic structure with a diffeomorphism. Crucially, the relative charge as well as the Komar charge both involve derivatives of the vector field $\xi$, which may be non-vanishing on $S$ even if $\xi$ itself is vanishing on $S$. This is precisely the case for surface boosts. The surface boost transformations are therefore represented non-trivially in the EH formulation of gravity, while they are trivially represented in the GR formulation.

To understand this difference we need to compute the algebra of charges, which we will denote by $\mathfrak{g}^S$. In the covariant phase space, this algebra is given by the Poisson brackets defined as $\{\CH^S[\xi],\CH^S[\zeta]\}=-\CL_\xi\ip\CL_\zeta\ip\Omega$. As expected for diffeomorphisms, computing this algebra for both the canonical GR and the EH formulations of gravity leads to
\be\label{general diff algebra}
\{\CH^S[\xi],\CH^S[\zeta]\}=\CH^S[\xi,\zeta]\,,
\ee
where $[\xi,\zeta]$ is the Lie bracket of vector fields. Stopping at this expression, one could erroneously conclude that canonical GR and EH gravity have the same corner symmetry algebra. This is however not the case, as revealed by a $2+2$ decomposition of the charges and the algebra \eqref{general diff algebra}. Indeed, the question is that of how exactly the diffeomorphism algebra \eqref{general diff algebra} is \textit{represented}.

As shown in \cite{Donnelly:2016auv} and recalled in Appendix \ref{2+2 Komar}, the algebra of the Komar charges obtained in EH gravity is given by the semi-direct product $\mathfrak{g}^S_\EH=\text{diff}(S)\ltimes\sll(2,\mathbb{R})^S_\per$. Here $\text{diff}(S)$ denotes the algebra of infinitesimal diffeomorphisms tangent to the boundary sphere $S$, generated by vector fields $\xi^a\pa_a$ with $a=1,2$ labelling the sphere coordinates, while $\sll(2,\mathbb{R})^S_\per$ denotes the algebra of \textit{surface boosts}. This latter is a normal subalgebra of $\mathfrak{g}^S_\EH$. The corresponding infinitesimal diffeomorphisms are vector fields $\xi^i\pa_i$, where $i=0,3$ label coordinates normal to $S$, whose components vanish on $S$ but possess non-zero normal derivatives, i.e. $\xi^i\stackrel{S}=0$ and $\nabla_i\xi^j \stackrel{S}\neq0$. They are germs of diffeomorphisms fixing $S$, hence surface boosts. A basis of this algebra of surface boosts can be labelled by  phase space functionals 
$J_a(x)$  which depend, in the EH formulation, on the conformal class of the normal metric\footnote{If we denote by $h_{ij}$, with $i,j\in\{0,3\}$ the components of the metric normal to $S$, by $\epsilon^{ij}$ the component of the Levi--Civita tensor, and by 
$\tau_a$ a $2\times2$ matrix representation of $\sll(2,\R)$ satisfying $\tau_a\tau_b=\tfrac{1}{4}\eta_{ab}+\tfrac{1}{2}\epsilon_{ab}{}^c\tau_c$, we define
\be
J_a(x)\coloneqq\sqrt{q}\,\f{h_{jk}\epsilon^{ki}}{\sqrt{|h|}}(\tau_a)_i{}^j.
\ee}.They satisfy the  commutation relations 
\be 
\{J_a(x),J_b(y)\}={\epsilon_{ab}}^c J_c(x)\delta^{(2)}(x,y),
\ee
where the Casimir element satisfies
\be
\f{1}{4}\det\big({q}(x)\big)=-J_0^2(x)+J_1^2(x)+J_2^2(x),
\ee
with $q_{ab}$ the sphere metric. The crucial difference between the GR and the EH formulations is that this surface boost algebra $\sll(2,\mathbb{R})^S_\per$ is trivially represented in GR, since the corresponding generators identically vanish there. As a result, the algebra of corner symmetries of EH gravity is larger than that of GR, which is simply given by $\mathfrak{g}^S_\GR=\text{diff}(S)$.

So far we have established that the two potentials for metric gravity, namely the canonical one $\Theta_\GR$ and the Einstein--Hilbert one $\Theta_\EH$, lead to inequivalent corner symmetry algebras. The boost subalgebra $\sll(2,\mathbb{R})^S_\per$ is trivially represented in canonical gravity with the potential $\Theta_\GR$. On the other hand, taking into account the presence of the corner symplectic potential in the Einstein--Hilbert formulation leads to a non-trivial representation of $\sll(2,\mathbb{R})^S_\per$.

With this in mind, we can now step back in order to explain the conceptual meaning of this result and its generalization to other formulations of gravity. This will bring forward the notion of trivial vs gauge redundancies, and make the case for edge modes.

\subsection{Corner symmetries and edge modes}

We now come to an important point about the role of corner degrees of freedom in gravity. From now on we assume that we are studying an entangling spacetime region $R$ which is sliced in terms of hypersurfaces $\Sigma$ that all hinge on a 2-dimensional corner $S$ (see Figure \ref{figure}). We denote by $\Theta$ the symplectic potential of a given formulation of gravity and $\Omega=\delta\Theta$ the corresponding symplectic form. As we have already seen on a concrete example, generically $\Omega$ will differ from $\Omega_\GR$ by a corner term. We want to understand the physics associated with different corner symplectic structures. Although the discussion in this section is general, in this work we explicitly study three different symplectic forms, namely $\Omega_{\GR}$, $\Omega_{\EH}$, and $\Omega_{\ECH}$, corresponding respectively to the canonical GR, Einstein--Hilbert, and Einstein--Cartan--Holst formulations of gravity. GR and EH have already been treated above as an introductory example, and in the companion paper \cite{Edge-Mode-II} we will study in details the case of Einstein--Cartan--Holst gravity (which we already briefly introduce in the next section).

It is essential to appreciate that each formulation possesses a different level of redundancy, or gauge symmetry. That is, each formulation realizes the same bulk theory in terms of different sets of variables. The canonical formulation refers to the choice of a foliation, i.e. a scalar field $T$ with slices $\Sigma_t=\{x\in R\,|\,T(x)=t\}$ and normal form $\un \propto \rd T$. The variables of the canonical formulation are the induced metric and the extrinsic curvature $(\tilde{g}_{ab},\tK^{ab})$, and the gauge redundancies are associated with the group $\mathcal{G}_\GR=\Sigma\mathrm{Diff}(R)$ of hypersurface-preserving diffemorphisms. Infinitesimally, these correspond to vector fields that satisfy $\tilde{g}_\mu{}^\nu\CL_\xi n_\nu=0$. In adapted coordinates $(T,x^a)$, where $x^a$ are coordinates on $\Sigma$, this means that the component $\xi^T$ of $\xi=\xi^T\partial_T+\xi^a\partial_a$ is independent of $x^a$, i.e. $\partial_a\xi^T=0$. In the Einstein--Hilbert formulation however, the variables are the spacetime metric components $g_{\mu\nu}$, and the gauge redundancies are the spacetime diffeomorphisms preserving $R$, that is $\mathcal{G}_\EH=\mathrm{Diff}(R)$. The variables of the Einstein--Cartan--Holst formulation of gravity, which will be reviewed in more details in the next section, are the frame fields $e_\alpha^I$. The corresponding gauge redundancies are the semi-direct product $\mathcal{G}_\ECH=\mathrm{Diff}(R)\ltimes\mathrm{SL}(2,\mathbb{C})^R$ of diffeomorphisms with local Lorentz transformations\footnote{We denote $G^X$, where $G$ is a group and $R$ a space, the set of maps $X\to G$.}.

All these formulations of gravity differ in the size of their gauge redundancies, and we can write that
\be 
\mathcal{G}_\GR\subset\mathcal{G}_\EH\subset\mathcal{G}_\ECH.
\ee
This ordering makes it clear that some formulations are more covariant\footnote{Here we use the term ``covariant'' in a cavalier manner to mean formulations with variables transforming non-trivially under the action of a bigger gauge group. For instance, the metric $g_{\mu\nu}$ is invariant under Lorentz transformations, so Lorentz transformations act \emph{trivially } on the metric. The frame field, on the other had, transforms non-trivially and therefore \emph{covariantly} under local Lorentz transformations. And even if both Lagrangians $L_{\EH}[g]$ and $L_{\ECH}[e]$ are invariant under Lorentz transformations, according to the terminology used here, the Einstein--Cartan--Holst Lagrangian has a bigger group of covariance.} than others, although one could also say that they are more redundant. We could also argue that the bigger group is always a redundancy from the point of view of the smaller theory, albeit a trivial one. For instance, the local Lorentz symmetry trivially leaves the metric invariant, and similarly, some diffeomorphisms can change the foliations and leave the corresponding canonical variables on a slice unchanged. These are the trivial redundancies of each formulation. This then raises a fundamental question: Is there any physical difference between these formulations of gravity? A common viewpoint is that, after all, the distinction between trivial redundancies and gauge redundancies is a matter of taste and does not change the classical physics. Since redundancies are unphysical, one should pick the formulation which has the least number of them. It is clear from our previous analysis that the GR formulation possesses less redundancies than the Einstein--Hilbert formulation. In that respect, canonical GR is a more minimal formulation of gravity. One can wonder whether it is possible to reduce further the gauge group of canonical GR without introducing some form of non-locality, and whether a minimal formulation exists for which even the diffeomorphism charges vanish. We leave this investigation for future work, noting that the fully gauged fixed formulation of gravity proposed in \cite{Grant:2009zz,Verlinde:1991iu}, or \cite{Sachs:1962zzb} in the case of null boundaries, could be such a minimal formulation.

 In any case, if gauge symmetry is mere redundancy, we should strive to write it in its minimal form. There is also, on the other hand, a sense in which covariant gravity, given by the Einstein--Hilbert formulation, and maybe first order gravity, given by the Einstein--Cartan (--Holst) formulation, lead to a more geometric, deeper formulation of the relativity principle. So which is which, and does it matter at all?

The answer to that question is a subtle and important one. One can argue that classically this does not really matter. However, at the quantum level, this is of crucial importance. The differences are physical, and there is a clear sense in which the theory with the bigger group provides a more extensive description of the quantum physical degrees of freedom. The key idea behind this was formulated in \cite{Donnelly:2016auv}. In the absence of a theory of quantum gravity, we can retreat for the analysis of this question to a semi-classical analysis. There, instead of studying quantum operators, quantum observables, quantum algebras, and their representations, we consider the classical phase space, the Poisson bracket, and the corresponding semi-classical algebras.

The main idea is that when considering bounded regions there is a subset of transformations, which are redundancies in the bulk, which becomes physical symmetries on the boundary. The main point is that different formulations of the theory (here, gravity) lead to different (inequivalent) representations of the corner symmetry group. This, in turn, means that these different formulations lead to inequivalent quantizations (i.e. different spectra for physical observables). What happens then is that the formulation with the smaller gauge redundancies (like canonical GR) represents some  corner symmetries trivially, while the formulation with the larger group represents non-trivially these corner symmetries. The reason behind this is that the formulations with larger symmetry groups, which are more covariant in the bulk, possess more boundary degrees of freedom. These degrees of freedom are distinguished by the action of the corner symmetry group. At the quantum level, these are the edge modes which appear as representation states for the corner symmetries.

Let us illustrate this in the case of diffeomorphism symmetry. We focus on the diffeomorphisms which preserve the entangling region. Infinitesimally, these correspond to vector fields $\xi$ whose pull-back on the hinging corner $S$ vanishes, i.e. $\xi\stackrel{S}=0$. Such diffeomorphisms form a canonical algebra, and as recalled in the previous section, given a gravitational symplectic potential $\Omega$, we can construct the Hamiltonian generator $\CH[\xi]$ associated with the infinitesimal diffeomorphism $\CL_\xi$ as
\be
-\CL_\xi\ipp\Omega =\delta\CH[\xi].
\ee
This Hamiltonian generator can then be decomposed as the sum of a bulk and corner generator
\be
\CH[\xi]=\CH^\Sigma[\xi]+\CH^S[\xi]. 
\ee
For a gauge redundancy, we then have that the bulk generator vanishes on-shell, i.e. $\CH^\Sigma[\xi]\approx0$. This means that the Hamiltonian generator is a pure corner term satisfying the corner algebra \eqref{general diff algebra}.

This defines the corner symmetry algebra, and the boundary generator $\CH^S[\xi]$ provides a representation of this symmetry. For a trivial redundancy however, we also have  that the corner generator itself vanishes. This gives the first clean distinction between gauge and trivial redundancies. Gauge redundancies correspond to bulk transformations which have a vanishing bulk Hamiltonian generator but a non-zero corner generator. Trivial redundancies are transformations which have a vanishing generator even when their parameter does not vanish on the corner.

Now, as we have also seen in the previous section, different formulations of gravity possess distinct corner symplectic potentials. This means that they also lead to distinct representations of the corner symmetry algebra. Even when the corner symmetry algebra is non-trivial, it is possible to chose the corner symplectic potential in such a way that the charge vanishes. In this case, the corner symmetry algebra is trivially represented, and this is the sign that we are in a formulation where the gauge redundancy has been trivialized. In such formulations, the corner symmetry algebra is not big enough to account for all the boundary degrees of freedom, and we cannot hope for a bulk reconstruction from the knowledge of the boundary charges. From this perspective, it is clear that one should look for formulations that have the biggest corner symmetry group, and not the smallest. That is, we should look for \textit{maximally extended theories}. The more extended the formulation, the bigger the corner algebra, the more we can reconstruct bulk degrees of freedom and dynamics from its boundary.

The bulk reconstruction follows from the fact that the conservation of boundary charges gives us an expression of the bulk constraints. Since the dynamics of gravity is entirely formulated in terms of constraints, we need the maximal amount of non-trivial corner symmetries in order to encapsulate all the dynamics. Another fundamental reason why we should look for maximally extended theories follows from our experience that the quantization of gauge theories starts from the quantization of boundary observables and their representations. Without the proper and complete set of non-trivial boundary observables, we do not have a proper handle on quantization.
 
The extension of a theory from a smaller corner symmetry group to a bigger one requires the addition of boundary degrees of freedom encoded into the choice of the symplectic corner potential. These are the elusive edge modes. Elusive because their presence is not mandatory if we just want to describe the classical bulk theory. However, they are necessary in order to achieve a bulk reconstruction and for a proper understanding of quantization. They are the reason for which we investigate here the different formulations of gravity and their various corner symmetry algebras. We want to understand more precisely how the nature and the size of the corner symmetry group depend on the chosen formulation. For instance, and as seen in the previous section, the Einstein--Hilbert formulation possess an extra canonical pair $(\sqrt{q}\,\tilde{s}_\mu,n^\mu)$ on the boundary. This extra canonical pair allows for the non-trivial representation of the surface diffeomorphism boost symmetry. Similarly, the Einstein--Cartan--Holst formulation contains, compared to Einstein--Hilbert, non-trivial charges for local Lorentz transformations, which encode additional information about the boundary frame. In summary, extended theories have a larger corner symmetry algebra, they possess more boundary degrees of freedom and activate more channels of bulk reconstruction. These are the reasons for which one should look for the maximally extended theory. It is then natural to wonder how to extend the theories, and how do we know that the maximal extension is reached? These are fundamental questions which will be addressed in future work. For the moment, we study the extensions that the Einstein--Hilbert and Einstein--Cartan--Holst formulations of gravity provide, relative to canonical GR, this latter being the minimal local representation of gravity at hand. As can be seen on Table \ref{table}, it is however clear that neither EH nor ECH represent the maximal extensions.

\subsection{Boundary conditions and Lagrangians}
We want to emphasize again that the discussion so far is about corner symmetries acting at the corner of spacetime on entangling spheres and that the corner symmetry group is independent of the choice of boundary conditions.
Boundary conditions are necessary when we want to extend the symmetry generator to a time development of the boundary symmetry group.
Lets call $\Delta=S\times I$ this time development.
In order to promote to boundary symmetry generators some of the corner symmetry generators, we need to impose boundary conditions on $\Delta$. Given the symplectic potential $\theta_L$ associated to the Lagrangian $L$ (see appendix \ref{Noethera} for the explicit construction), a general boundary condition $B$ is of the form 
$\theta_L \stackrel{\Delta} = \rd \vartheta_B-\delta \ell_B   $
see \cite{Iyer:1994ys, Harlow:2019yfa}.
We say that a boundary condition $C$ is \emph{canonical } if there exists a Lagrangian  $L_C$ such that the boundary condition is simply $\theta_{L_C}\stackrel{\Delta} =0$.

We have seen---in \eqref{correspondence} and equations  (\ref{theta-boundary}, \ref{thetab})---that a boundary Lagrangian $\ell$ admits a corner symplectic potential $\vartheta_{\ell}$.
This means that a general boundary condition is canonical when $\vartheta_B =\vartheta_{\ell_B}$.
When this is the case, we can define  $L_B := L + \rd \ell_B$.
The equation \eqref{theta-boundary} simply means that 
$\theta_{L_B}= \theta_L + \delta \ell_B - \rd\vartheta_B $ so that the boundary condition $B$ is equivalent to $\theta_B \stackrel{\Delta} =0$ hence is canonical.
This shows that the choice of Lagrangian can also be related to a choice of canonical boundary conditions.
The detail interplay between the corner symmetry group (independent of boundary conditions but dependent on Lagrangians) and the boundary symmetry group that depends only on the choice of boundary condition will be discussed more thoroughly in a separate publication.

\section{Tetrad gravity}
\label{sec:tetrad}

We now turn to the tetrad formulation of gravity, and we study the Einstein--Cartan and Einstein--Cartan--Holst Lagrangians and symplectic potentials. 
We give the construction of the boundary charges and  shows that the corner symmetry is $\text{diff}(S)\ltimes\sll(2,\mathbb{C})^S$. Our ultimate goal is to show that the symplectic potential of tetrad gravity decomposes, as for metric gravity \eqref{EH-GR potentials relation}, into the bulk piece $\Theta_\GR$ plus a corner piece and that the additional corner piece is responsible for the corner symmetry extension $\text{diff}(S)\to\text{diff}(S)\ltimes\sll(2,\mathbb{C})^S$. For tetrad gravity the bulk + boundary decomposition  of the symplectic potential is more subtle than the metric case and requires more work. We defer its detailed analysis to the companion paper \cite{Edge-Mode-II}. In preparation for the tetrad gravity case, let us treat BF theory first.

\subsection{BF theory}

We are going to introduce tetrad gravity as a topological BF theory where the field $B$ satisfies the so-called simplicity conditions. Properly understanding, both in the bulk and on the boundary, the role and the meaning of these simplicity conditions in the covariant phase space formalism, first requires to properly understand BF theory itself. In this section we will therefore recall some basic features of BF theory. 

In the 4-dimensional case which we are studying, BF theory is constructed with a Lorentz tensor 2-form $B^{IJ}$ and a Lorentz connection 1-form $\omega^{IJ}$ with curvature $F^{IJ}$. The Lagrangian is
\be
L_\text{BF}[B,\omega]=  \f{1}{2} B_{IJ}\wedge F^{IJ}[\omega].
\ee
Its variation is
\be\label{variation BF Lagrangian}
\delta L_\text{BF}[B,\omega]=\f{1}{2}\left(\delta B_{IJ}\wedge F^{IJ}+\delta\omega^{IJ}\wedge T_{IJ}\right) +\f{1}{2} \rd\left(B_{IJ}\wedge\delta\omega^{IJ}\right),
\ee
with $T_{IJ}\coloneqq\rd_\omega B_{IJ}$. From this one can see that the symplectic current is $\theta=\tfrac{1}{2}B_{IJ}\wedge\delta\omega^{IJ}$, while the bulk equations of motion are the flatness and Gauss equations
\be
F^{IJ}\approx0,\q T_{IJ}\approx0.
\ee
As it is well-known, this theory is topological i.e. possesses no local degrees of freedom. This can be understood by counting the degrees of freedom. In the canonical formulation, the phase space variables are the spatial components of $B_{IJ}$ and $\omega^{IJ}$, which are a total of $18+18=36$ variables. The theory only has first class constraints, which are the pullbacks to $\Sigma$ of the equations of motion. Taking into account the Bianchi identity, there is a total of $18-6$ curvature constraint plus $6$ torsion constraints, and all are first class constraints. The counting indeed gives zero phase space degrees of freedom. The topological nature of the theory is also reflected in the gauge symmetries of the theory, which in addition to diffeomorphisms and internal Lorentz transformations contain the so-called translations (or shifts). We will return to the analysis of these gauge transformations shortly.

\subsubsection{Symplectic potential and charges}
\label{sec:BF charges}

The symplectic potential of BF theory is simply given by
\be\label{bare BF potential}
\Theta_\text{BF}=\f{1}{2} \int_\Sigma B_{IJ}\wedge\delta\omega^{IJ}.
\ee
With this potential, it is then straightforward to build the symplectic structure and to contract it with the infinitesimal gauge transformations of the theory in order to build the Hamiltonian generators and the boundary charges \cite{Mondragon:2004nw}. We recall these results in the next section because they apply straightforwardly to tetrad Einstein--Cartan--Holst gravity once we impose simplicity constraints expressing $B^{IJ}$ in terms of the coframe $e^I$.
The reader can also check Appendix \ref{Noethera} for more details on the canonical charges in the covariant phase space formalism.

We now study the Hamiltonian charges. Some of the results presented here will translate immediately to the case of tetrad gravity. The symplectic structure of BF theory is given by
\be
\Omega_\text{BF}=\f{1}{2} \int_\Sigma\delta B_{IJ}\wedge\delta\omega^{IJ}.
\ee
With this symplectic structure we can proceed as in the case of metric gravity and study the Hamiltonian generators and charges for gauge transformations. In the case of topological BF theory, there are three types of such transformations: Lorentz gauge transformations $\delta_\alpha$ labelled by a Lie algebra-valued function $\alpha^{IJ}$, diffeomorphisms $\CL_\xi$ labelled by a vector field $\xi=\xi^\mu\partial_\mu$, and translations labelled by a Lie algebra-valued 1-form $\phi^{IJ}$.

Let us first focus on Lorentz transformations. They act on the fields as
\be\label{Lorentz transformations}
\delta_\alpha B^{IJ}=[B,\alpha ]^{IJ}
,\q\delta_\alpha\omega^{IJ}=\rd_\omega\alpha^{IJ},
\ee
and this action satisfies $[\delta_\alpha,\delta_\beta]=\delta_{[\alpha,\beta]}$, where $[\alpha,\beta]^{IJ} $ denotes the Lie algebra commutator. We can contract this transformation with the symplectic structure to find
\be
-\delta_\alpha\ipp\Omega_\text{BF}=\frac{1}{2}\int_\Sigma\left(\delta B_{IJ}\wedge\delta_\alpha\omega^{IJ}-\delta_\alpha B_{IJ}\wedge\delta\omega^{IJ}\right)=\delta\CH_\text{BF}[\alpha],
\ee
with
\be
\CH_\text{BF}[\alpha]\coloneqq\f{1}{2}\int_\Sigma B_{IJ}\wedge\rd_\omega\alpha^{IJ}.
\ee
It is important to appreciate that this equality is valid when $\alpha$ is  field-independent, that is when $\delta\alpha=0$. This Hamiltonian can be split through integration by parts into bulk and corner components
\be
\CH_\text{BF}[\alpha]=\CH_\text{BF}^\Sigma[\alpha]+\CH_\text{BF}^S[\alpha],
\ee
where the bulk component is the constraint and the corner piece is the charge
\be
\CH_\text{BF}^\Sigma[\alpha]\coloneqq-\f{1}{2}\int_\Sigma\alpha^{IJ}T_{IJ}\approx0,\q\CH_\text{BF}^S[\alpha]\coloneqq\f{1}{2}\int_S\alpha^{IJ}B_{IJ}.
\ee
We therefore get that the generator of Lorentz transformations is a boundary term defined by the boundary value of $B$. These generators satisfy the canonical algebra
\be\label{BF Lorentz algebra}
\{\CH^S_\text{BF}[\alpha],\CH^S_\text{BF}[\beta]\}=\CH^S_\text{BF}[\alpha,\beta].
\ee
We can conclude from this analysis that Lorentz transformations that vanish on $S$ are gauge transformations, while transformations with a non-zero parameter $\alpha$ on $S$ are {\it symmetry} transformations labelling different boundary states at the quantum level. The associated symmetry group is the loop group $\SL(2,\mathbb{C})^S$, and the boundary $B$ field is the canonical generator of the loop algebra $\sll(2,\mathbb{C})^S$. Given $(x,y)\in S$ we have that
\be
\{B_{IJ}(x),B_{KL}(y)\}=\left(\delta_{JK}B_{IL}-\delta_{IK}B_{JL}-\delta_{JL}B_{IK}+\delta_{IL}B_{JK}\right)(x)\delta^{(2)}(x,y).
\ee
The fact that the boundary $B$ field is non-commutative \cite{Cattaneo:2016zsq} has many implication for the quantization of the theory as we will see.

We now turn to the study of diffeomorphisms. They act on differential forms as usual as the Lie derivative $\CL_\xi=\xi\ip(\rd\,\cdot)+\rd(\xi\ip\cdot)$, where $\ip$ denotes the spacetime contraction of vector fields and forms (we use the same symbol for field-space contraction as well). The action of diffeomorphisms can be straightforwardly evaluated to find
\be\label{diffBF}
-\CL_\xi\ipp\Omega_\text{BF}=\delta\left(\f{1}{2}\int_\Sigma B_{IJ}\wedge\CL_\xi\omega^{IJ} \right)-\f{1}{2}\int_S\xi\ip\left(B_{IJ}\wedge\delta\omega^{IJ}\right).
\ee
This action is Hamiltonian if and only if the diffeomorphism preserves the boundary sphere, i.e. when the pull-back of $\xi$ on $S$ is a vector tangent to $S$, which we  assume. In this case the last term in \eqref{diffBF} vanishes. Using that 
\be\label{Lie of omega}
\CL_\xi\omega=\xi\ip F+\rd_\omega(\xi\ip\omega),
\ee
the Hamiltonian
\be
\CH_\text{BF}[\xi]=\f{1}{2}\int_\Sigma B_{IJ}\wedge\CL_\xi\omega^{IJ}
\ee
can be separated into bulk and corner components
\be
\CH_\text{BF}^\Sigma[\xi]=\f{1}{2}\int_\Sigma\left(\xi\ip B_{IJ}\wedge F^{IJ}+\xi\ip\omega^{IJ}T_{IJ}\right)\approx0,\q\CH_\text{BF}^S[\xi]=\f{1}{2}\int_S \xi\ip\omega_{IJ}B^{IJ}.
\ee
Note that here we have used the same notation $\CH^S_\text{BF}$ for the generators, which are distinguished by their arguments: $(\alpha,\beta)$ stand for Lie algebra elements and are used for Lorentz transformations, while $(\xi,\zeta)$ are vector fields and are used for diffeomorphisms. 

We therefore get a non-vanishing charge only when $\xi$ does not vanish on $S$. This means that the corner symmetry subalgebra due to diffeomorphisms is simply $\text{diff}(S)$. This fact will obviously remain true when going to tetrad gravity by imposing the simplicity condition relating $B$ to the coframe $e$. One can therefore already anticipate that the corner symmetry algebra of tetrad gravity will differ from that of Einstein--Hilbert metric gravity. Indeed, as we reviewed in Section \ref{sec:metric}, the latter contains the semi-direct product of $\text{diff}(S)$ with the boost algebra $\sll(2,\mathbb R)^S_\per$ generated by vector fields that vanish on $S$ but have a non-vanishing normal derivative.

We can now look at the translations, which are specific to topological BF theory. Given a Lie algebra-valued 1-form $\phi^{IJ}$, the translations act on the fields as
\be
\delta_\phi B^{IJ}=\rd_\omega\phi^{IJ},\q\delta_\phi\omega^{IJ}=0.
\ee
One can see that these are indeed symmetries of the theory by plugging them in the variation \eqref{variation BF Lagrangian}, since then, up to a boundary term obtained with an integration by parts, one obtains the Bianchi identity $\rd_\omega F^{IJ}\equiv0$. Contracting a translation with the symplectic structure leads to
\be
-\delta_\phi\ipp\Omega_\text{BF}=-\int_\Sigma\delta_\phi B_{IJ}\wedge\delta\omega^{IJ}=\delta\CH_\text{BF}[\phi],
\ee
with
$\CH_\text{BF}[\phi]=\CH_\text{BF}^\Sigma[\phi]+\CH_\text{BF}^S[\phi]$ given by  the decomposition
\be
\CH_\text{BF}^\Sigma[\phi]\coloneqq-\int_\Sigma\phi_{IJ}\wedge F^{IJ}\approx0,\q\CH_\text{BF}^S[\phi]\coloneqq-\int_S\phi_{IJ}\wedge\omega^{IJ}
\ee
into bulk constraint and corner charge. These BF translations do not preserve the simplicity constraints and therefore are not relevant for the analysis of tetrad gravity. However, it is well-known that the diffeomorphisms, Lorentz transformations, and translations are not independent. Indeed, the diffeomorphisms can be expressed, up to the equations of motion, in terms of field-dependent Lorentz transformations and translations. This can easily be seen by noticing that the Lie derivative acting on $B$ or $\omega$ can be written as
\be
\CL_\xi=\xi\ip(\text{EOMs})+\delta^\text{translation}_{\xi\ip B}+\delta^\text{Lorentz}_{\xi\ip\omega},
\ee
with the corresponding equations of motion EOMs depending on whether the diffeomorphism acts on $B$ or $\omega$.

\subsection{Einstein--Cartan--Holst formulation}
\label{subsec:ECH}

The tetrad formulation of gravity involves an $\mathbb{R}^4$-valued 1-form, or coframe field $e^I=\rd x^\mu e_\mu{}^I$, with inverse $\hat{e}_I=\hat{e}_I{}^\mu\partial_\mu$. This coframe field is related to the spacetime metric via $g_{\mu\nu}=e_\mu{}^I e_\nu{}^J\eta_{IJ}$, where $\eta_{IJ}=\text{diag}(-1,1,1,1)$ is a kinematical Lorentz metric. Frames and coframes are related by the inversion formula $\hat{e}_I{}^\mu=g^{\mu\nu}\eta_{IJ}e_\nu{}^J$. The coframe locally defines a $\text{GL}(4,\mathbb{R})$, while the metric is invariant under the local Lorentz group and belongs to the coset space $\text{GL}(4,\mathbb{R})/\text{SO}(3,1)$. In addition to this coframe, the tetrad formulation involves a Lorentz connection 1-form $\omega^{IJ}$ with curvature 2-form $F^{IJ}$. Like in the previous section, we will first review the action and its equations of motion, before briefly looking at the symplectic potential and the boundary charges. The detailed analysis of the symplectic potential and its bulk-boundary decomposition is performed in \cite{Edge-Mode-II}, and here we only summarize the results.

In terms of the coframe and the Lorentz connection as independent dynamical variables, the bulk Lagrangian for tetrad gravity which we are going to study in the rest of this paper is the first order Einstein--Cartan--Holst (ECH) Lagrangian
\be\label{EC Lagrangian}
L_\ECH[e,\omega]= \f{1}{2} E_{IJ}[e]\wedge F^{IJ}[\omega],\q E_{IJ}[e]\coloneqq(\as+\beta)(e\wedge e)_{IJ}.
\ee
It is obtained by taking the Lagrangian of topological BF theory studied in the previous section and imposing $B^{IJ}\stackrel{!}{=}E^{IJ}[e]$. This condition is equivalent to the simplicity constraints
\be
\as B^{IJ}-\beta B^{IJ}\q\mathrm{is\ simple},\q\Leftrightarrow\q
\big(\as B^{IJ}-\beta B^{IJ}\big)\wedge\big(\as B^{KL}-\beta B^{KL}\big)=V\epsilon^{IJKL},
\ee
where $V$ is the 4-volume, meaning that $(\as -\beta)B^{IJ}$ can be written as the wedge product of coframe fields. The duality map acting on the Lie algebra is defined as $2(\as M)_{IJ}={\epsilon_{IJ}}^{KL}M_{KL}$. The parameter $\gamma=\beta^{-1} $ is the so-called Barbero--Immirzi parameter \cite{Barbero:1994ap,Immirzi:1996di}, and its presence is the reason why this Lagrangian has the name Holst \cite{Holst:1995pc} appended to Einstein--Cartan. According to what is shown on Table \ref{table}, we will see in \cite{Edge-Mode-II} with the bulk + boundary decomposition that the presence of this Holst term accounts for the $\su(2)$ part of the corner symmetry algebra.

The variation of this Lagrangian gives the bulk equations of motion and the symplectic potential. This takes the form
\be\label{variation EC action}
\delta L_\ECH[e,\omega]=\delta e_I\wedge G^I+\delta\omega^{IJ}\wedge T_{IJ}+\f{1}{2}\rd(E_{IJ} \wedge\delta\omega^{IJ}),
\ee
where
\be\label{EC EOMs}
G^I\coloneqq(\as+\beta)F^{IJ}\wedge e_J\approx0,\q T_{IJ}\coloneqq\f{1}{2}\rd_\omega E_{IJ}=\f{1}{2}(\as+\beta)\big(\rd_\omega e_{[I}\wedge e_{J]}\big)\approx0,
\ee
with $G^I$ the Einstein tensor in tetrad variables. When the coframe is invertible, the second equation implies the vanishing of the torsion $T^I\coloneqq\rd_\omega e^I$, and upon imposing this torsion-free condition the first equation becomes Einstein's equations of motion. This is because, given an invertible coframe, there is a unique torsionless connection $\omega=\gamma[e]$ such that
\be
\rd_\gamma e^I=\rd e^I+\gamma^I{}_J\wedge e^J=0.
\ee
The solution to this equation is given by the Koszul formula with
\be
\gamma_\mu^{IJ}[e]=\left(\delta_\mu^\alpha\big(\hat{e}^{I\beta }\delta^J_K-\hat{e}^{J\beta }\delta^I_K\big)-\hat{e}^{I\alpha }\hat{e}^{J\beta }e_{\mu K}\right)\partial_{[\alpha}e_{\beta]}{}^K.
\ee
Inserting this expression in the initial Lagrangian leads to the second order form of the Einstein--Cartan Lagrangian, and the Holst term proportional to $\beta$ vanishes identically by virtue of the Bianchi identity. An expression which will be  useful   is the contraction of the torsionless connection with a vector field $\xi$. Denoting $\xi^I\coloneqq\xi\ip e^I$ the associated internal vector and $\underline\xi=e^I\xi_I=\xi_\mu\rd x^\mu$ the associated 1-form, we get 
\be\label{gammaip}
\xi\ip\gamma^{IJ}=\hat{e}^{[I}\ip\CL_\xi e^{J]}-\hat{e}^I\ip\hat{e}^J\ip\rd\underline\xi.
\ee

The two sets of first order equations \eqref{EC EOMs} satisfy two Bianchi identities. The first one is the expression of local Lorentz invariance, i.e. the invariance $\delta_\alpha L=0$ of the Lagrangian under the Lorentz transformations \eqref{Lorentz transformations} and $\delta_\alpha e^I=-{\alpha^I}_Je^J$, and reads
\be 
\rd_\omega T_{IJ}=e_{[I}\wedge G_{J]}.
\ee 
The second Bianchi identity is the expression of diffeomorphism invariance, i.e. the consequence of the transformation law $\CL_\xi L=\rd(\xi\ip L)$ for the Lagrangian under the action of a diffeomorphism. Given a vector field $\xi$ and its contraction $\xi_I=\xi\ip e_I$, this second Bianchi identity takes the form
\be
\xi_I \rd_\omega G^I=\xi\ip T_I\wedge G^I+\xi\ip F^{IJ}\wedge T_{IJ}.
\ee
In particular, when the torsion vanishes these Bianchi identities become the usual symmetry and conservation conditions on the Einstein tensor, i.e.
\be
e_{[I}\wedge G_{J]}\simeq0,\q\rd_\omega G^I\simeq0,
\ee
where we denote by $\simeq$ the torsionless condition $\rd_\omega e^I\simeq0$. Since in what follows we will alternatively impose either all the equations of motion, or only the half corresponding to the torsionless condition, we will separate these two cases with the full on-shell $\approx$ and half on-shell $\simeq$ equalities.

\subsubsection{Symplectic potential and charges}
\label{sec:EC potential}

From the boundary term in \eqref{variation EC action} we can read off the Einstein--Cartan--Holst symplectic potential
\be\label{EC initial potential}
\Theta_\ECH=\f{1}{2} \int_\Sigma E_{IJ}[e]\wedge\delta\omega^{IJ}.
\ee
Using this   symplectic potential, the analysis of the Hamiltonian charges follows verbatim that of the charges in BF theory, with the only difference that the translations do not exist in tetrad gravity, and that the $B$ field has to be replaced by $E[e]$. The Hamiltonian charges for gauge and diffeomorphism are \cite{Mondragon:2004nw,Engle:2010kt, Bodendorfer:2013sja,Corichi:2013zza}
\be
\CH_\ECH[\alpha]=\f{1}{2}\int_\Sigma E_{IJ}\wedge\rd_\omega\alpha^{IJ},\q\CH_\ECH[\xi]=\f{1}{2}\int_\Sigma E_{IJ}\wedge\CL_\xi\omega^{IJ}.
\ee
The Gauss and diffeomorphism constraints can be written as the conditions
\be 
\CH_\ECH[\alpha]\approx\CH_\ECH[\alpha'],\q\CH_\ECH[\xi]\approx\CH_\ECH[\xi']\q\alpha\stackrel{S}=\alpha',\q\xi\stackrel{S}=\xi',
\ee
where $\alpha,\alpha'$ are Lie algebra-valued functions on $\Sigma$ that agree on $S$, and $\xi,\xi'$ are tangent vectors on $\Sigma$ that agree on $S$ \cite{Balachandran:1995qa}. The common sphere value are the corner Hamiltonian charges
\be
\CH_\ECH^S[\alpha]\coloneqq\f{1}{2}\int_S\alpha^{IJ}E_{IJ},\q\CH_\ECH^S[\xi]\coloneqq\f{1}{2}\int_S\xi\ip\gamma_{IJ}E^{IJ}.
\ee
These Hamiltonians are functionals of the boundary frame. Using the explicit contraction \eqref{gammaip} of the connection with a vector field, we can write the boundary diffeomorphism charge as the sum of two terms
\be 
\CH_\ECH^S[\xi]=\f{1}{2}\int_S E^{IJ}(\hat{e}_{I}\ip\CL_\xi e_{J})-\f{1}{2}\int_S E^{IJ} \hat{e}_I\ip\hat{e}_J\ip\rd\underline\xi.
\ee  
The second term is the Komar charge. The first term can be given a canonical interpretation as a relative charge\footnote{The notion of {\it relative charge} has been introduced in \eqref{relative-charge} and also used in \eqref{relative-charges2}. It is given by the difference between Hamiltonian charges associated to two different formulations of gravity, or directly as a canonical charge coming from the relative pre-symplectic potential, namely the difference between the pre-symplectic potentials associated to the two  different formulations (see \cite{Edge-Mode-II} for additional examples).} \cite{DePaoli:2018erh,Oliveri:2019gvm,Edge-Mode-II}.

The charge algebra simply reflects the commutator relations $[\CL_\xi,\CL_\zeta]=\CL_{[\xi,\zeta]}$, $[\CL_\xi,\delta_\alpha]=\delta_{\CL_\xi\alpha}$, and $[\delta_\alpha,\delta_\beta]= \delta_{[\alpha,\beta]}$. For tangential diffeomorphisms and Lorentz transformations we get
\be
\{\CH^S_\ECH[\xi],\CH^S_\ECH[\zeta]\}=\CH^S_\ECH[\xi,\zeta],\q\{\CH^S_\ECH[\alpha],\CH^S_\ECH[\beta]\}=\CH^S_\ECH[\alpha,\beta],
\ee
and the mixed bracket is given by
\be
\{\CH^S_\ECH[\xi],\CH^S_\ECH[\alpha]\}=\CH^S_\ECH[\CL_\xi\alpha].
\ee
Note that, as in the $\BF$ case above, we distinguish generators of different transformations   by their arguments.
The parameter $\CL_\xi\alpha$ is a Lorentz transformation, and reflects the fact that we have the semi-direct structure
\be
\text{diff}(S)\ltimes\sll(2,\mathbb{C})^S
\ee
as the charge algebra between tangent diffeomorphisms and Lorentz transformations. The local corner charges are explicitly given by 
\be
E^{IJ}\stackrel{S}=\f{1}{2}\epsilon^{ab}E_{ab}^{IJ},\q D_a\stackrel{S}=\gamma_a^{IJ}E_{IJ},
\ee 
with $a,b$ denoting indices tangent to $S$. We see that the generator of Lorentz transformations is a boundary term given by the boundary value of the geometrical flux $E_{IJ}[e]$. This is an important fact which demonstrates a key difference with the metric formulation of gravity for which boundary Lorentz transformations are pure gauge. This means that tetrad gravity possesses additional boundary degrees of freedom compared to metric gravity. These extra boundary degrees of freedom are edge modes, and they play a key role in the quantization of the theory. The generators satisfy the local current algebra\footnote{ The ultralocal   subalgebra  $\text{diff}(S)\ltimes\su(2)^S$  was first presented in \cite{Engle:2010kt}.
The Lorentz algebra was  considered in \cite{Bodendorfer:2013sja} for arbitrary dimension. They follows straightforwardly from the more general BF analysis.}
\begin{subequations}
\be
\{D_a(x),E_{IJ}(y)\}&=\pa_a\delta^{(2)}(x,y)E_{IJ}(x),\\
\{D_a(x),D_b(y)\}&=\pa_a\delta^{(2)}(x,y)D_b(x)-\pa_b\delta^{(2)}(x,y)D_a(y),\\
\{E_{IJ}(x),E_{KL}(y)\}&=\left(\delta_{JK}E_{IL}-\delta_{IK}E_{JL}-\delta_{JL}E_{IK}+\delta_{IL}E_{JK}\right)(x)\delta^{(2)}(x,y).
\ee
\end{subequations}
Importantly, one should notice that the corner symmetry algebra $\text{diff}(S)\ltimes\sll(2,\mathbb{C})^S$ is the same in Einstein--Cartan--Holst gravity and in topological BF theory. At the classical level they provide different realizations of the same commutation relations. At the quantum level they provide  \emph{different representations} of the corner symmetries. The main difference stems from the simplicity constraints, which have a very simple expression on a slice $\Sigma$, thanks to the presence of the internal normal $n^I$. The bulk simplicity constraints read 
\be\la{Bsimp}
(\as E)_{IJ}n^J\stackrel{\Sigma}=\beta E_{IJ} n^J.
\ee
In the Hamiltonian analysis, evolution of the above primary simplicity constraints in the bulk leads to  a set of conjugated secondary constraints \cite{BarroseSa:2000vx,Alexandrov:2006wt,Alexandrov:2011ab}. Now, as
both sides of the equality \eqref{Bsimp} are commuting variable, the primary bulk simplicity constraints \eqref{Bsimp} are first class {\it with themselves}, but they form a second class system with the secondary  bulk simplicity constraints. Imposition of this second class system of constraints leads to ambiguities in the definition of a Lorentz connection configuration
variable  in the bulk \cite{Alexandrov:2001wt,Alexandrov:2011ab}.
 In the covariant phase space formalism, on the other hand, one does not need to worry about secondary  bulk simplicity constraints, as the formalism is by definition on-shell in the bulk and the primary constraints are automatically preserved in time. 
 In this case then, the only set of 
 corner simplicity constraints one needs to impose seemingly takes the same form as  \eqref{Bsimp} for forms pulled back on $S$. However, since the corner variable are now non-commuting, the corner  simplicity constraint are also second class, but in this case with themselves. Therefore, as thoroughly explained in \cite{Edge-Mode-II}, also in the covariant phase space formalism there is a fundamental ambiguity in dealing with simplicity constraints, although of a different nature than in the Hamiltonian analysis. 
As we will see \cite{Edge-Mode-II}, this discrepancy between bulk and corner simplicity constraints, which means that we cannot simply treat the corner variable as a continuity of the bulk ones, is one the main reason  behind the necessity to introduced edge modes.
And they will be the main subject of paper \cite{Edge-Mode-III} in this series.

A particularly interesting restriction is when the time gauge is imposed in the Einstein--Cartan--Holst formulation. The choice of time gauge fixes $n^I=(1,0,0,0)$, and restricts the Lorentz transformations to preserve only an $\SU(2)$ subgroup. What is now clear is that the time gauge not only fixes the bulk gauge symmetry, but it also kills the boundary boost charges. This shows one important and subtle point of confusion in the analysis of gauge theories, which is that some gauge fixing, which affect boundary modes, can lead to a theory with less boundary degrees of freedom than the ones which are not gauged fixed. This is not the case for a differential gauge fixing such as the Lorentz gauge, but it is the case for gauge fixings such as the unitary gauge or the ones containing additional restrictions that fix uniquely the boundary frames such as in \cite{Gomes:2019xto}. In the case of gravity, when we chose the time gauge we restrict the boundary Lorentz symmetry to a boundary $\SU(2)$ symmetry. The generator of rotations is given, in the time gauge, by the LQG flux $E^i=\beta(e\times e)^i/2$, where the cross-product is $(e\times e)^i\coloneqq{\epsilon^i}_{jk}e^j\wedge e^k$. The corresponding degrees of freedom associated with this charge are constitutive to the definition of the quantum theory. The non-commutativity of rotation charges means that $\beta$ corresponds to the area gap at the quantum level. In particular one sees that when $\beta=0$ the rotational charges vanish and the LQG degrees of freedom are pure gauge, thus one obtains a different canonical and therefore quantum theory. In addition to these rotational, loopy degrees of freedom, tetrad gravity also exhibits boost degrees of freedom which are not accessible in the metric formulation of gravity nor in current LQG representations due to the imposition of the time gauge. This is essentially what is summarized on Table \ref{table}. The goal of this series of papers  is, in part, to reveal these boost degrees of freedom. They are encoded in the corner symplectic structure \eqref{decomposed ECH}, which we will derive in \cite{Edge-Mode-II} and study in depth in \cite{Edge-Mode-III}. In addition, in \cite{Edge-Mode-II} we show that a careful study of the corner symplectic structure \eqref{decomposed ECH} reveals the presence of a (tangent) $\sll(2,\mathbb{R})$ algebra, as indicated on Table \ref{table}. Since this derivation takes the paper in another direction, we choose this point to stop and conclude.

\subsubsection{Bulk-boundary decomposition}

We now give a description of the  bulk + boundary decomposition of the ECH potential.
This will then enable us to study in depth the corner symplectic potential, the boundary simplicity constraints, and the quantization of the frame field. This is the content of the follow-up papers in this series.
 The detailed derivation of the bulk + boundary decomposition of the potential, is deferred to the companion paper \cite{Edge-Mode-II}. Here we only summarize some  key results which reveal, as expected, that the bulk piece is the same universal contribution mentioned above, namely the GR bulk potential (written in tetrad variables), and that, as expected, the only difference between the metric and tetrad formulations of gravity lies in the form of the corner potential.

Given a space-like slice $\Sigma$ with normal form $\un$ and normal vector $\hat{n}$, we can introduce an internal normal $n^I=\hat{n}\ip e^I$ such that
\be
\un=e^In_I,\q\hat{n}=\hat{e}_In^I,\q n^In_I=n^\mu n_\mu=-1.
\ee
With this one can define the induced coframe as
\be
\tilde{e}^I_\mu\coloneqq e^I_\mu+n_\mu n^I.
\ee
This form is both \textit{tangential} in the sense $\tilde{e}^In_I=0$, and \textit{horizontal} in the sense $\hat{n}\ip\tilde{e}^I=n^\mu\tilde{e}_\mu^I=0$. It enables us to write the induced metric on $\Sigma$ as
\be
\tilde{g}_{\mu\nu}\coloneqq\tilde{e}_\mu^I\tilde{e}_\nu^J\eta_{IJ}=g_{\mu\nu}+n_\mu n_\nu.
\ee
Then, let us define the extrinsic curvature 1-form on $\Sigma$ as
\be 
\tK^I\coloneqq\rd_\omega n^I+\un\wedge\big(\hat{n}\ip\rd_\omega n^I\big).
\ee
This form is also both \textit{tangential} and \textit{horizontal}, since $\tK^In_I=0$ and $\hat{n}\ip\tK =0$. With these ingredients, one finds that on-shell of the torsionless condition  the potential \eqref{EC initial potential} becomes \cite{Edge-Mode-II}
\be
\Theta_\ECH\simeq\int_\Sigma\tilde{P}_I\wedge\delta\te^I-\delta\left(\f{1}{2}\int_\Sigma\tilde{P}_I\wedge\te^I\right)+\int_S\left(\tE_I\delta n^I-\f{\beta}{2}\te_I\wedge\delta\te^I\right),
\ee
where\footnote{We define the cross product as $(M\times N)^I\coloneqq{\epsilon^I}_{JKL}(M^J\wedge N^K)n^L$.} $\tilde{P}^I\coloneqq(\tilde{K}\times\tilde{e})^I$ and $\tE^I\coloneqq\frac12 (\tilde{e}\times\tilde{e})^I$. The bulk terms in this expression for the potential are nothing but
\be
\Theta_\GR=\int_\Sigma\tilde{P}_I\wedge\delta\te^I,\q\int_\Sigma\teps\tK=\f{1}{2}\int_\Sigma\tilde{P}_I\wedge\te^I.
\ee
For the symplectic structure this implies
\be\label{decomposed ECH}
\Omega_\ECH\simeq\Omega_\GR+\int_S\left(\delta\tE_I\wedge \delta n^I-\f{\beta}{2}\delta\te_I\wedge\delta\te^I\right),
\ee
which, as announced, is the decomposition of the ECH symplectic structure into the bulk GR piece and a corner piece.

One can see that the corner symplectic structure $\Omega_\ECH^S$ carries the two extra canonical pairs $(\tE_I,n^I)$ and $(\te^I_a,\te^I_b)$. The first canonical pair is the contribution of the Einstein--Cartan part of the Lagrangian. It has appeared previously in \cite{Peld_n_1994}, and is only non-trivial when the internal normal $n^I$ is included in the phase space and not fixed by a gauge choice. The role of this normal has been acknowledged (both in the bulk and on the boundary) in many occurrences. It is required for example in order to write down covariant boundary terms for the variational principle \cite{0264-9381-4-5-011,bianchi2012horizon,Bodendorfer:2013hla}. This normal also plays a prominent role in extensions of LQG beyond the time gauge \cite{Alexandrov:2002br,Alexandrov:2007pq,Alexandrov:2008da,Alexandrov:2011ab}. It is also central in proposals for the simplicial dynamics of LQG \cite{Wieland_2014,Wieland_2017} (see also \cite{Wieland:2017zkf} for the null case). Finally, it is used in extensions of LQG (and its underlying classical structure) to higher dimensions \cite{Bodendorfer_2013,Bodendorfer:2013jba}.

The corner symplectic structure also carries a $\beta$ contribution coming from the Holst term. In the time gauge where the normal is fixed to $n^I=\delta^I_0$, this follows simply from the fact that the Holst term in the bulk can be written as a boundary term using the identity\footnote{Here $i\in1,2,3$ are indices in the $\su(2)$ subalgebra of $\sll(2,\mathbb{C})$ which survives the fixing of the normal.} $2\det(e^i_a)e^a_i\delta\gamma^i_a[e]
=\rd(\delta e^i\wedge e_i)$, where $\gamma^i_a[e]$ is the torsionless spin connection \cite{Thiemann:2001yy,Liu:2009em}. This has also been used in studies of isolated horizons in LQG \cite{Engle:2010kt}.
Moreover, it is the central piece that was used in \cite{Freidel:2015gpa, Freidel:2016bxd, Freidel:2019ees} to propose a new quantization of gravity that includes the boundary metric, the momenta and the corner diffeomorphism charges in the 
observables that can be quantized. When the internal normal $n^I$ is arbitrary, the corner symplectic potential written above with its two contributions has appeared in this form in \cite{Bodendorfer:2013jba}.

The decomposition \eqref{decomposed ECH} and the corner symplectic structure will be the main focus of \cite{Edge-Mode-II,Edge-Mode-III}, where we use it to study the corner metric and its algebra, as well as the boundary simplicity constraints and their proper imposition and quantization. Finally, notice that this decomposition also extends to the decomposition of the diffeomorphism charges, which reads
\be
\CH_\ECH[\xi]= \CH_\GR[\xi] + \CH_{\ECH/\GR}[\xi],
\ee
where the charges inferred from the GR and ECH symplectic structures are
\be\label{two diffeo charges}
\CH_\GR[\xi]\simeq\int_S\tilde{P}_I (\xi\ip\te^I),
\q
\CH_\ECH[\xi]\simeq\f{1}{2}\int_S(\xi\ip\omega)_{IJ}E^{IJ},
\ee
while the relative canonical charge is
\be
\CH_{\ECH/\GR}[\xi]=\int_S\left(\CL_\xi n^I \tE_I  + \f{\beta}2 \CL_\xi\te_I\wedge \te^I\right).
\ee
The form of the GR diffeomorphism charge $\CH_\GR[\xi]$ given in \eqref{two diffeo charges} is simply a rewriting of the Brown--York charge \eqref{BY charge}, and has been studied in \cite{Freidel:2019ees,Freidel:2019ofr} as the starting point of a new quantization of diffeomorphism symmetry.

\section{Conclusion}

A careful analysis of the covariant phase space of metric gravity has shown how different Lagrangians differing by a choice of boundary term reveal different components of the corner symmetry algebra. More precisely, the symplectic potential can always be decomposed into a common bulk term, parametrized by the ADM canonical conjugate variables, plus a corner term that varies according to the choice of boundary Lagrangian. The corner potential encodes degrees of freedom living on the boundary and associated to non-trivial charges. Moreover, different formulations of gravity extend the structure of the bulk gauge group, and this has a direct reflection in the extension of the corner symmetry group.

In this way, our analysis reveals how the \textit{ambiguity} in the  definition of the symplectic structure due to the presence  of boundary terms can be systematically solved by unambiguously relating a corner potential to a given boundary Lagrangian. These boundary degrees of freedom acquire a precise physical meaning as they provide a non-trivial representation of a new component of the corner symmetry algebra. Moreover, these degrees of freedom constitute the \textit{raison d'${\hat{e}}$tre}  of the edge modes that need to be introduced to extend the boundary phase space.  As we will see in the subsequent papers, these edge modes play a dual role. First, as was already discussed in previous work, they enable to restore gauge-invariance. Second, and on a deeper level, they can be used to restore  time conservation of the charges while relaxing the boundary conditions as much as possible. This mechanism will be  explained in details in the subsequent papers of this series, but it is already clear that boundary terms do not represent an unnecessary complication of the covariant phase space formalism: Instead, they can be exploited to provide an organizing principle thanks to which different notions of quasi-local physical charges can be given an Hamiltonian interpretation. As it will become clear along the way, this is also the principle that unveils the physical nature of edge modes and reveal   different fragments of the treasure map.

\section*{Acknowledgement}

Research at Perimeter Institute is supported in part by the Government of Canada through the Department of Innovation, Science and Economic Development Canada and by the Province of Ontario through the Ministry of Colleges and Universities.
We would like to thank Alejandro Perez for many  discussions, inspirations, encouragements and an early participation in some of the key insights.

\appendix

%
 \section{Fields and jets}
\label{appendix1}

It is well known that the space of fields denoted $\CF$ can be viewed as the space 
\be
\CF:= \Gamma(E,M)
\ee of sections of  a given vector bundle $p:E\to M$,  over the manifold $M$, referred to as the spacetime. 
Locally,  we have the identification $U_E\simeq U_M \times U_F$, for open sets $U_E\in E$, $U_M\in M$,  and $U_F=\pi^{-1}(U_M)$ is the fiber neighborhood.
 An element $\Phi \in \CF$  can be viewed as  a map 
$\Phi : M \to F$ where $F$ is the  fiber.
Once we chose local coordinates $(x^\mu,\varphi^A)$ on $U_M\times U_F$,  we can view fields as maps  $\Phi: U_M \to U_F $ given by $ x\to\varphi^A(x)$.

The symmetry group $\CG={\mathrm{Aut}(E)}$  is given by the set of automorphisms of $E$. 
By definition $\CG$ is the subset of $\mathrm{Diff}(E)$ which projects onto $\mathrm{Diff}(M)$.
The infinitesimal group of automorphism of $E$ comprises  of \emph{projectable} vector fields: A vector fields $\xi \in \mathfrak{X}(E)$ is projectable if it can be written in local coordinates as
\be
\xi =\xi^\mu(x) \pa_\mu + \xi^A(x,\varphi)\pa_A. 
\ee 
And local change of coordinates preserving the bundle structure are given by invertible maps $(x,\varphi) \to (x',\varphi')$ such that  
$
x'^\mu = F^\mu(x)$ and $\varphi'^A = F^A(x,\varphi). 
$

\subsection{Observables}

By definition  a  field $\Phi \in \CF $ defines a map 
\be
\Phi: M \to E, \qquad \Phi  \circ p= {\mathrm{Id}}.
\ee
This map can be used to pull back forms $\Phi^*: \Omega(E) \to \Omega(M)$ and define local field observables  on $M$   as $ \int_M \Phi^*(\omega)$ for a top form $\omega \in \Omega(E)$.
This set of local observables is too restrictive to introduce interesting Lagrangian, it contains only 
integral of functional of $\Phi$ with no derivative.
In order to construct more interesting observables we need to extend the bundle $E$ to the jet bundle $JE$.

$J^kE$, the space of k-jets  of local sections of $E$, is defined  as the set of equivalence classes of pairs $(x, s)$,  $x\in M$ and $ s $ a  local section of $E$, defined on a neighborhood 
of $x$;  with equivalence $(x, s) \sim (x', s')$  if and only if $x = x'$ 
and all the derivatives of $s$  and $s'$ up to and including order $k$ are equal at $x$. $J^kE$ has, in the obvious way, the structure of a smooth manifold.
There are canonical projections $\pi_k^l: J^kE\to J^lE$  whenever $k\geq l\geq0$.
In particular since $J^0E= E$ we will be particularly interested in 
\be
\pi_k : J^kE \to E, \qquad p_k : J^kE \to M\,,
\ee
with $p_k =\pi_k\circ p$ and $p_0=p$. The space of $\infty$-jets of local sections, which is denoted by $J E$  is, as a set, the projective limit of the system 
$(J^kE,\pi^l_k)$. 

A field $\Phi \in \CF$ naturally gives a section of the jet bundle 
\be
\Phi: J E \to M , \qquad \Phi \circ p_\infty = \mathrm{Id}. 
\ee
And we can use this to define the space of local observables
by pulling back smooth\footnote{A Function $P$ on $JE$ is said to be $C^\infty$ 
if for each $s \in JE$ there exists $k\in\mathbb{N}$, $U_k$ a neighbourhood of 
the projected $k$-jet $\pi^k_\infty(s)\in J^kE$ and $P_k \in C^\infty(J^kE)$ such that 
$F|_{\pi^k_\infty(U_k)} = F \circ \pi^k_\infty$.} forms  on $JE$ and integrating them on $M$.

A set of local coordinate of $JE$ is given by  $ (x,[s])=(x^\mu,\varphi^A,\varphi_\mu^A, \varphi^A_{\mu\nu}, \cdots)$.
It is convenient to introduce the multi-index $\bmu=(\mu_1,\cdots ,\mu_n)$, denote 
$|\bmu|=n$ and $\varphi^A_{\bmu}= \varphi^A_{(\mu_1,\cdots ,\mu_n)}$ and $\pa_{\bmu}=\pa_{\mu_1} \cdots \pa_{\mu_n}$. $x^\mu$ is a coordiante on the base manifold $M$ while $\varphi^A_{\bmu}$ are the fiber's coordinates.
A field $\Phi$, which is a section of $E$, is extended (tautologically) to give a section of  the jet bundle. In local coordinates this gives
\be
\Phi^*(\varphi^A)=\Phi^A(x),\qquad \Phi^*(\varphi^A_{\boldsymbol\mu}) =\pa_{\boldsymbol\mu} \Phi^A(x). 
\ee 
More generally, given a smooth function $P$ on $JE$, 
$P(x,\varphi_\bmu^A)$, we define the local  field observables 
\be\label{Obs}
O_P(\Phi):= \Phi^*P, \q \mathrm{ or} \q  O_P(\Phi) [x]= P(x,\pa_{\bmu}\Phi^A(x)).
\ee
If one wants to construct field theory Lagrangian one needs to allow not only the pull-back of 
functions but also the pull-back of forms. 
A Lagrangian is then an element of $\Omega_n(JE)$, a $d$-form on the jet bundle with $d=\mathrm{dim}(M)$, and the action is obtained by integrating its pull-back over $M$ 
\be 
S =\int_M \Phi^* L .
\ee
In full generality, we can generalize the formula \eqref{Obs}, and  associate a field observable  to forms of any degree. 
Given $\omega \in \Omega_{\bullet}(JE)$, the space of local forms on field-space is given by the pull-back $\Omega^{\mathrm{loc}}(\CF, M):= \Phi^*( \Omega(JE))$.
A local field observable in $\Omega^{\mathrm{loc}}(\CF, M)$ is given by 
\be 
O_\omega(\Phi) :=  \Phi^*(\omega).
\ee 
This means that we can organize the field-space forms in terms of a \emph{bidegree}.
$O_\omega  \in \Omega^{\mathrm{loc}}_{\bullet \bullet}(\CF) $ is said to be of degree 
$(\mathsf{a},a)$ if the spacetime degree of $O_\omega$ is $a$ and its total degree is 
$\mathsf{a}+a$. $\mathsf{a}$ is said to be the field-space degree of the form $O_\omega$. 

The reason it is necessary to introduce the notion of jet bundle is that it allows us to  bypass the unnecessary complications of defining field differentials on infinite dimensional functional space
(see  \cite{Brouder_2018} and references therein, for a comprehensive review).
The Jet bundle allows us to define the variational derivative as the pull back of usual   derivatives. 

\subsection{Cartan calculus}

Given a field $\Phi_A(x)$ we can define its derivative and repackage the knowledge of its derivative in terms of the Cartan differential $\rd\Phi  =\rd x^\mu \pa_\mu\Phi $, where $\rd x^\mu$
is a basis of one form on $M$. The Cartan differential satisfies the defining relation $\rd^2=0$.
More formally  we can consider the de Rham algebra $(\Omega(M), \wedge )$ on $E$.
It is a $\Z$-graded algebra, with $\Z$-grading given by  the form degree
$\alpha\wedge \beta= (-1)^{ab} \beta\wedge \alpha$ for  form $(\alpha,\beta) \in \Omega_a(M)\times \Omega_b(M)$.
A graded derivative $D$ of degree $d$ is a map $D:\Omega_\bullet(M)\to \Omega_{\bullet+d}(M)$ that satisfy
\be
[D,\alpha\wedge]= (D\alpha)\wedge. 
\ee 
The commutators are graded commutators: Given two operators of degree $a$ and $b$ their commutator is  
$[A,B]:=AB-(-1)^{ab}BA$.
This algebra carries 3-graded derivative: The differential $d$ of degree $-1$, The Lie derivative  ${\CL}_\xi$ of degree $0$ and  the interior product $\xi \ip$  of degree $-1$.
The last two  are associated with a vector field $\xi \in \mathfrak{X}(M)$.
They satisfy the 6 Cartan axioms:
\bea\label{ax1}
[\rd, \rd ] =0, \qquad [\xi \ip,\zeta \ip ]&=&0,\qquad [\rd, \xi \ip ]=\CL_\xi,\\
{}[\rd, \CL_\xi ] =0,\qquad [\CL_\xi , \zeta \ip ]&=&[\xi,\zeta] \ip , \qquad  [\CL_\xi ,\CL_{\zeta} ]=
\CL_{[\xi,\zeta]}\,,\label{ax2}
\eea
where the commutators are graded commutators.
The first three axioms stipulates that $\rd$ is a differential and that interior product anti-commutes while the last one is the magic Cartan formula which defines the Lie derivatives as a graded commutator of $\rd$ and interior product.
The last three control the commutation of the Lie derivative.
These are not independent axioms since they follow simply from the use of the graded Jacobi identity
$ [[A,B],C]= [A,[B,C]]+ (-1)^{cb}[[A,C],B]$. For instance 
\bea
0=[[\rd,\rd],\xi\ip]= [\rd,[\rd,\xi\ip]] - [[\rd,\xi\ip],\rd]= 2[\rd, \CL_\xi ].
\eea
And similarly for the other identities.
 
An algebra $(A,\wedge,\ip, \rd)$ satisfying the axioms (\ref{ax1},\ref{ax2}) is called a \emph{Cartan differential graded algebra} (Cartan-dga).
Note that the fact that $\xi \ip,\rd$ and $\CL_\xi$ are derivation means that we effectively have three additional Cartan axioms controlling the commutation of the wedge product
\be
[\rd, \alpha \wedge ] =\rd\alpha\wedge,\qquad [\xi \ip , \alpha \wedge ] =(\xi \ip \alpha)\wedge  , \qquad  [\CL_\xi ,\alpha \wedge ] =( \CL_\xi\alpha)\wedge . 
\ee

\subsection{Generalized Cartan calculus}

We present here the variational calculus, more details about it can be found in 
\cite{Anderson, Barnich:2001jy, Olver-Lie, Krasil_shchik_2011, Compere:2018aar}.
The field differential is a differential that satisfies the Cartan relation and acts as a derivative 
\be 
\delta^2 =0\,.
\ee
To define $\delta$ we start with its action on the field coordinates $\Phi(x)$
and generalizes its action on a field functional $O_P(\Phi)$ by linearity and using the Leibniz rule.
For instance the action of $\delta $ on the field components $\pa_\bmu \Phi^A$ is determined by the fact that the operation $\Phi_A\to \pa_\bmu \Phi^A$ is a linear operation which simply commute with the 
variational differential and therefore 
$\delta \pa_\bmu \Phi^A(x):= \pa_{\bmu} \delta \Phi_A(x)$.
More formally,  for a 
local field observable $O_P(\Phi)= \Phi^*P$ with $P\in \Omega(JE)$
we have 
\be
\delta O_P(\Phi)[x]   = \sum_{|\bmu|=0}^\infty \pa_\bmu \delta\Phi^A(x) \wedge  
\pa_A^\bmu O_{P}(\Phi)[x], \qquad 
\pa_A^\bmu O_{P}(\Phi) := O_{\frac{\pa P}{\pa \varphi_\bmu^A}}(\Phi).
\ee
In the covariant phase space formalism, one therefore needs to introduce the  the bigraded space of forms on field-space $\Omega_{\bullet,\bullet}(M,\CF)$.
One of the grading is the de Rham grading that gives the degree of the form in spacetime (the numbers of $\rd$),
while the other grading is the field-space grading that assign the degree of the form in field 
space (the number of $\delta$'s). A General form  $O_P$ is of bidegree $(p,\mathsf{p})$ 
with $p$  de-Rham degree and $\mathsf{p}$ the field degree.
For instance $ \delta \phi^A \wedge \phi^*(P_{A\mu\nu}) \rd x^\mu\wedge\rd x^\nu$,
with $P_{A\mu\nu} \in C^\infty(JE)$  is  a form of bidegree $(2,1)$.
The wedge product of bigraded forms is defined to be 
\be
(P)\qquad  O_P\wedge O_Q= (-1)^{pq+\mathsf{p}\mathsf{q}} O_Q\wedge O_P.
\ee
where $O_P$ is of bidegree $(p,\mathsf{p})$ and  $O_Q$ is of bidegree $(q,\mathsf{q})$.
We call this convention the physicists ($P$) convention of bigrading, while mathematicians usually use the total degree as a grading. We stick to the physicists convention in our work.

The variational Cartan calculus  also requires the definition of the notion of a field-space vector field, which is an element of $\mathfrak{X}(\CF) := \Omega_{0,1}(M,\CF)^*$.
A field-space vector field $X\in \mathfrak{X}(\CF)$ is constructed from the knowledge 
of a generalized vector field $\bar{X}$ on $E$. A generalized vector field on $E$ is a vertical vector field  
$\bar{X}:= \bar{X}^A( x, \varphi^A_\bmu) \pa_A$ with coefficient being jet functions, $\bar{X}^A\in C^\infty(JE)$.
The associated field-space vector field $X\in \mathfrak{X}(\CF)$ can be denoted 
\be
X =\int_M X^A(\Phi)[y] \frac{\delta }{\delta \Phi^A(y)},\qquad
X^A(\Phi) =\Phi^* \bar{X}^A.
\ee
It is  defined by its action on the fundamental variables and the fact that it commute with the 
spacetime derivative
\be
X [\Phi^A] 
=
X^A(\Phi),
\qquad 
X [\pa_{\bmu}\Phi^A] 
=\pa_{\bmu}X^A [\Phi].
\ee
Its action on a general local field functional  $O_P\in C(\CF)=O_{0,0}(M,\CF)$ follow from the Leibniz rule,
\be
X[O_PO_Q]= X[O_P] O_Q + O_PX[O_Q].
\ee
Explicitely this means that  given $P\in C^{\infty}(\CF)$, we have  
\be
 X[O_P] =\sum_{|\bmu|=0}^\infty \pa_\bmu X^A \pa_A^\bmu O_P.
\ee
The field-space vector fields form a Lie algebra with bracket denoted $[X,Y]$.

Given $X\in \mathfrak{\CF}$ we can define the field interior product $X\ip$,
it is a derivation of bi-degree $(0,-1)$ and it simply acts on the fundamental forms 
as $X\ip(\delta\pa_{\bmu}\Phi^A(x)) = \pa_\bmu X^A(x)$.
We can also introduce the field Lie derivative 
$L_X$ and  construct the variational calculus where $(\Omega_{p,\bullet}(M,\CF),\delta, \ip, \wedge)$ form a Cartan-dga:
\bea\label{axd1}
[\delta, \delta ] =0, \qquad [X \ip,Y \ip ]&=&0,\qquad [\delta,X \ip ]=L_X,\\
{}[\delta, L_X ] =0,\qquad [L_X , Y \ip ]&=&[X,Y] \ip , \qquad  [L_X ,L_Y ]=
L_{[X,Y]}.\label{axd2}
\eea
To complete the bi-graded Cartan calculus we have to specify how the 
variational calculus interacts with the de-Rham calculus.
We demand that variational differentials of degree $(0,1)$ commute with de-Rham differential of degree $(1,0)$ and that de-Rham interior product of degree $(-1,0)$ also commutes with the field interior products
\be 
[\rd,\delta]=0,\qquad [X\ip, \xi\ip]=0.
\ee
The only things left to specify is the cross-commutator between differential and interior products.
Given $\xi \in \mathfrak{X}(M)$ and $X\in \mathfrak{X}(\CF)$ we have 
\be\label{crossc}
 [\rd, X\ip] = 0, \qquad
[\delta , \xi \ip]= (\delta \xi) \ip. 
\ee
The first commutator together with the other Cartan axioms  imply also that the de-Rham Lie derivative commutes with the  field contraction: 
\be 
[\CL_\xi, X\ip]=0,\qquad [L_X,\rd]=0.
\ee
The second commutator  involves the contraction of a mixed object,  the variational vector field 
 $(\delta  \xi^\mu) \pa_\mu   \in \Omega_{-1,1}(M,\CF)$. It vanishes for diffeomorphisms that are field independent.
 Using this commutator and graded Jacobi we can establish that 
 \be 
 [L_X,\xi\ip]= (L_X\xi) \ip,\qquad
 [L_X,\CL_\xi]= \CL_{L_X\xi}.
 \ee

\subsection{Noether analysis}\label{Noethera}

We now have all the tools to perform the Noether analysis \cite{Crnkovic:1987tz,Crnkovic:1986ex,Zuckerman,gotay1998momentum}. Starting from a Lagrangian $d$-form $L\in \Omega_{d,0}(M,\CF)$, one constructs its equations of motion $ E_L \in  \Omega_{d,1}(M,\CF)$, and its pre-symplectic potential $\theta_L \in \Omega_{d-1,1}(M,\CF)$.
That is we construct a linear map
\bea 
 \Omega_{d,0}(M,\CF) &\to&  \Omega_{d,1}(M,\CF) \oplus \Omega_{d-1,1}(M,\CF) \cr
 L&\to& (E_L,\theta_L)
\eea
Which defines \emph{unambiguously}, given $L$, the equations of motion and the symplectic potential.

In order to describe this map we introduce the Euler differential operators, whose action on field space functionals is
\be 
 \Pi^\bmu_A[O_P]  :=  \sum_{|\bnu|=0}^\infty (-1)^{|\bnu|} \frac{(|\bmu|+|\bnu|)!}{|\bmu|!|\bnu|!}  \pa_{\bnu} {\pa_A^{\bmu\bnu}}O_P,
\ee
for a function $P\in C(JE)$.
This differential operators  are such that 
\be
\sum_{|\bmu|=0}^\infty (\pa_\bmu X^A) \pa_A^\bmu O_P =  \sum_{|\bmu|=0}^\infty
D_\bmu\left( X^A \Pi_A^\bmu[O_P] \right),\qquad \Pi_A^{\bmu}(\pa_\alpha O_P) = \Pi_A^{\bmu/\alpha}[O_P].
\ee
\begin{dfn}\label{defet}
Using these we can define $E_L$ and $\theta_L$ as 
\be 
E_L:= \delta \Phi^A  \pi_A[L],\qquad
\theta_L:= \sum_{|\bmu|=0}^\infty \pa_{\bmu} \left( \delta \Phi^A  \Pi_A^{\bmu\alpha}[L_\alpha] \right),
\ee
where $L_\alpha :=\pa_\alpha \ip L$ is a codimension-$1$ form.
\end{dfn}
An important property of the differential calculus we have just described is that 
the map $L\to (E_L,\theta_L)$ is un-ambiguously defined once a choice of coordinates on field space is made.
This is a key component of our construction  that explains why different Lagrangians possess different symplectic structures. The map $L \to E_L$ defining the equation of motions is well-known.
The map $L \to \theta_L$ has almost never been used  in the literature on covariant calculus. 
One exception is  the work of Lee and Wald \cite{Lee:1990nz} who proposed a un-ambiguous prescription for  $\theta_L$ which coincide with ours.
In the case where the Lagrangian is first order, that is depends only on $\phi^A$ and $\pa_\alpha \Phi^A$, the prescription for the symplectic potential is simply
\be
E_L = \left(
\frac{\pa L}{ \pa \phi^A }-
\pa_\alpha \left(\frac{\pa L}{ \pa_{\alpha}\phi^A}\right)   
\right) \epsilon, \qquad \theta_L := 
\delta \phi^A 
\left(\frac{\pa L}{\pa \pa_{\alpha}\phi^A }\right) \epsilon_\alpha. 
\ee
where $\epsilon$ is the volume form and $\epsilon_\alpha =\pa_\alpha \ip \epsilon $ denote the codimension one volume forms.
The fact that there is a well defined prescription for first order Lagrangian is well-know \cite{Julia:2002df}. 
We simply extend the definition to an arbitrary Lagrangian.

$\theta_L\in \Omega_{d-1,1}(M,\CF)$, which is unequivocally  assigned to $L$,  is  both a codimension-$1$ form in spacetime and a 1-form in field-space.
The fact that it  can be interpreted as the symplectic potential associated with the Lagrangian $L$ follows from the following  key properties.
\begin{prop}
$E_L$ and $\theta_L$ satisfy the key relation
\be\label{main-var}
 \delta L = E_L + \rd \theta_L. 
 \ee
 Moreover, given $\ell \in \Omega^{d-1}(M)$ we have 
\be\label{theta-boundary}
E(\rd \ell) =0, \qquad \theta_{\rd \ell} = \delta \ell - \rd \vartheta_{\ell},
\ee
where \be \label{thetab}
\vartheta_{\ell}= \sum_{|\bmu|=0}^\infty \frac{|\bmu|+1}{|\bmu|+2} \pa_{\bmu} \left( \delta \Phi^A  \Pi_A^{\bmu\alpha}[\ell_\alpha] \right).
\ee
Finally, we have that the prescription is covariant.
Under a diffeomorphism $\phi$, we have
\be
\phi^*E_L = E_{\phi^*L},\qquad 
\phi^*\theta_L =\theta_{\phi^*L}, 
\ee
where $\phi^*$ denotes the pull-back.
\end{prop}

The fact that this proposition is satisfied  follows from the generalized Cartan calculus rules that we have introduced and the Anderson homotopy operators \cite{Anderson}.
A detailed  proof and more complete discussion of this construction will appear in \cite{GeneralizedL}.
We see in particular that a shift of the Lagrangian by a total differential  implies that $L$ and $L+\rd \ell$ pre-symplectic potential differ by a corner term 
$\rd\vartheta_{\ell}$.

 The Euler operator  $E_L $ defines the equations of motion, while  $\theta_L$  defines the symplectic structure. 
 Taking the field differential of $\theta_L$ and   
  integrating over a slice $\Sigma$ produce the symplectic structure
\be
\Omega_L:=\int_\Sigma\delta\theta_L, \qquad \Omega_L(X,Y)= Y\ip X\ip \Omega_L.
\ee
A field variation  $\delta_\alpha=  \alpha^a R_a{}^A \frac{\delta}{\delta \Phi^A}$ associated with a field-independent parameter $\alpha$ can then be viewed as a derivation on field-space. 
It is a symmetry if there exists a spacetime vector field $\alpha^\sharp\in \mathfrak{X}(M)$
such that $\delta_\alpha L= \rd (\alpha^\sharp \ip L)$.
In this case we can show that the Noether current associated with $L$ is 
\be
J_L[\alpha]:= \delta_\alpha \ip \theta_L - \alpha^\sharp \ip L\,.
\ee
This is conserved on-shell since $\rd J_L[\alpha] = \delta_\alpha \ip E_L$ and one defines the Noether charges 
\be 
\CH_L[\alpha] :=\int_\Sigma J_{L}[\alpha].
\ee
One can establish that 
\be
\delta_\alpha \Omega_L +\delta \CH_L[\alpha]  =  \int_{S} (\alpha^\sharp \ip \theta_L) + \int_{\Sigma} (\alpha^\sharp \ip E_L).
\ee
This means that the transformation $\delta_\alpha$ is canonical with  Hamiltonian 
$\CH_L[\alpha] $ if it satisfy the boundary condition $\alpha^\sharp \ip \theta_L\stackrel{S}=0$. 
One also sees that if $\alpha^\sharp \ip E_L\stackrel{\Sigma}=0$, the transformation is canonical
even when we do not impose the on-shell conditions. In this case the symmetry transformation is \emph{kinematical}.
Note that under a shift $L\to L+\rd \ell$ we have that
\be
J_{L +\rd \ell}[\alpha]= J_{L}[\alpha] - \rd J_\ell[\alpha],
\qquad 
J_\ell[\alpha] = \delta_\alpha \ip \vartheta_{\ell} - \alpha^\sharp \ip \ell. 
\ee
We see that a shift of the Lagrangian by a total derivative implies a shift of the charges by a corner charge and a shift of the admissible boundary conditions
\be
 \CH_{L+\rd\ell} [\alpha]=\CH_L[\alpha]-\int_{\pa_\Sigma} J_{\ell}[\alpha],
 \qquad \alpha^\sharp \ip\theta_{L+\rd \ell} \stackrel{S}=0\,\,  \to \,\, \alpha^\sharp \ip\theta_{L}\stackrel{S}= \delta (\alpha^\sharp \ip\ell) - \alpha^\sharp \ip \rd \vartheta_\ell.
\ee

The Poisson bracket of Noether  Hamiltonians is then defined to be
\be
\{\CH_L[\alpha],\CH_L[\beta]\}=\delta_\alpha\CH_L[\beta]=-\Omega_L(\delta_\alpha,\delta_\beta).
\ee
A fundamental result in the covariant phase space formalism is that we can split $\CH[\alpha]$ as the sum
\be
\CH_L[\alpha]=\CH_L^\Sigma[\alpha]+\CH_L^S[\alpha],
\ee
where the bulk piece is vanishing on-shell, i.e. $\CH_L^\Sigma[\alpha]\approx 0$ for a gauge transformation, and where the boundary piece $\CH_L[\alpha]\approx\CH^S[\alpha]$ is traditionally called the Hamiltonian charge.

Note that for a \emph{kinematical} symmetry we have that the pre-symplectic potential transforms as
\be
L_{\delta_\alpha} \theta_L=\delta(\alpha^\sharp \ip L),
\ee
The case where  $\alpha^\sharp \ip L=0$ is particularly appealing since then the Hamiltonian charge is simply $\CH_L[\alpha] = \int_\Sigma (\delta_\alpha\ipp\theta_L)$.

\section{Einstein--Hilbert pre-symplectic potential}
\label{appendix:EH potential}

We recall that $n^\mu$ is the unit normal vector to the slice $\Sigma$, such that $g_{\mu\nu}n^\mu n^\nu=n^\mu n_\mu=-1$, and that $\tilde{g}_{\mu\nu}=g_{\mu\nu}+n_\mu n_\nu$ is the induced metric on $\Sigma$. For any vector $v^\mu$ we have that
\be
\delta(\nabla_\mu v^\mu)=\nabla_\mu\delta v^\mu+\delta\Gamma^\mu_{\mu\nu}v^\nu,\q\delta(\nabla_\mu v_\nu)=\nabla_\mu\delta v_\nu-\delta\Gamma^\rho_{\mu\nu}v_\rho,
\ee
so
\be
\delta(\nabla^\mu v_\mu)=\delta(g^{\mu\nu}\nabla_\mu v_\nu)=\delta g^{\mu\nu}\nabla_\mu v_\nu+g^{\mu\nu}\delta(\nabla_\mu v_\nu)=\delta g^{\mu\nu}\nabla_\mu v_\nu+\nabla^\mu\delta v_\mu-g^{\mu\nu}\delta\Gamma^\rho_{\mu\nu}v_\rho.
\ee
This enables us to write 
\be
2\delta(\nabla_\mu v^\mu)
&=\delta(\nabla_\mu v^\mu+\nabla^\mu v_\mu)\nn\\
&=\nabla_\mu\delta v^\mu+\nabla^\mu\delta v_\mu+\delta g^{\mu\nu}\nabla_\mu v_\nu-v_\mu(g^{\alpha\beta}\delta\Gamma^\mu_{\alpha \beta}-g^{\mu \beta}\delta\Gamma^\alpha_{\alpha\beta})\nn\\
&= 2\nabla_\mu\delta v^\mu_\per+\delta g^{\mu\nu}\nabla_\mu v_\nu-v_\mu\theta^\mu_\EH,
\ee
where
\be
\delta v^\mu_\per\coloneqq\f{1}{2}(\delta v^\mu+g^{\mu\nu}\delta v_\nu).
\ee
This definition is such that 
\be
n_\mu \delta v_\perp^\mu =\f12 \delta (n_\mu v^\mu). 
\ee
We can now introduce the derivative operator defined by the action
\be
\tilde{\nabla}_\mu v_\nu=\tilde{g}_\mu{}^\alpha\tilde{g}_\nu{}^\beta\nabla_\alpha v_\beta,\q\tilde{\nabla}_\mu v^\nu=\tilde{g}_\mu{}^\alpha\tilde{g}_\beta{}^\nu\nabla_\alpha v^\beta.
\ee
In particular, we have
\be\label{induced derivative n}
\tilde{K}_{\mu\nu}=\tilde{\nabla}_\mu n_\nu=\nabla_\mu n_\nu+n_\mu\tilde{a}_\nu,
\ee
where $\tilde{a}_\mu=n^\alpha\nabla_\alpha n_\mu$ is the acceleration. For any vector $v^\mu$ such that $v^\mu n_\mu=0$ we have
\be
\tilde{\nabla}_\mu v^\mu=\tilde{g}_\alpha{}^\beta\nabla_\beta v^\alpha=\nabla_\mu v^\mu+n_\mu n^\alpha\nabla_\alpha v^\mu=\nabla_\mu v^\mu-n^\alpha\nabla_\alpha n_\mu v^\mu=\nabla_\mu v^\mu-\tilde{a}_\mu v^\mu.
\ee
In particular, we  have that
\be
\nabla_\mu\delta n^\mu_\per=\tilde{\nabla}_\mu\delta n^\mu_\per+\tilde{a}_\mu\delta n^\mu_\per,
\ee
where the vector $\delta n^\mu_\per$ lives on $\Sigma$ since $\delta n^\mu_\per n_\mu=0$. Now, using
\be
\delta n^\mu=\delta g^{\mu\nu}n_\nu+g^{\mu\nu}\delta n_\nu
\ee
we also get that
\be
2\tilde{a}_\mu\delta n^\mu_\per=\tilde{a}_\mu\delta n^\mu+\tilde{a}^\mu\delta n_\mu=n_\mu\tilde{a}_\nu\delta g^{\mu\nu}+2\tilde{a}^\mu\delta n_\mu.
\ee
We can therefore write
\be
2\nabla_\mu\delta n^\mu_\per=2\tilde{\nabla}_\mu\delta n^\mu_\per+n_\mu\tilde{a}_\nu\delta g^{\mu\nu}+2\tilde{a}^\mu\delta n_\mu.
\ee
Finally, we get that
\be\label{decomposed EH potential}
2n_\mu\theta^\mu_\EH
&=-2\delta(\nabla_\mu n^\mu)+2\nabla_\mu\delta n^\mu_\per+\delta g^{\mu\nu}\nabla_\mu n_\nu\nn\\
&=-2\delta(\nabla_\mu n^\mu)+2\tilde{\nabla}_\mu\delta n^\mu_\per+\delta g^{\mu\nu}\tilde{K}_{\mu\nu}+2\tilde{a}^\mu\delta n_\mu\nn\\
&=-2\delta\tilde{K}+2\tilde{\nabla}_\mu\delta n^\mu_\per+\delta\tilde{g}^{\mu\nu}\tilde{K}_{\mu\nu}+2\tilde{a}^\mu\delta n_\mu,
\ee
where we have used the fact that $\delta g^{\mu\nu}\tilde{K}_{\mu\nu}=\delta\tilde{g}^{\mu\nu}\tilde{K}_{\mu\nu}$. One last step finally leads to
\be\label{mainEH-GR}
\boxed{\quad\tilde{\eps}\,n_\mu\theta^\mu_\EH=\f{1}{2}\tilde{P}^{\mu\nu} \delta\tilde{g}_{\mu\nu}-\left(\delta(\tilde{\eps}\tilde{K})-\teps\,\tilde{a}^\mu\delta n_\mu\right)+\tilde{\eps}\,\tilde{\nabla}_\mu\delta n^\mu_\per,\quad}
\ee
with
\be
\tilde{P}^{\mu\nu} \coloneqq\tilde{\eps}(\tilde{K}\tilde{g}^{\mu\nu}-\tilde{K}^{\mu\nu}).
\ee
We see in this formula that for a general variation we have an additional term $\tilde{a}^\mu\delta n_\mu$. One can restrict the set of variations to preserve the foliation, which means that we impose the restriction $\delta n_\mu\propto n_\mu$. Since $\tilde{a}^\mu n_\mu=0$, the last but one term in the potential drops in this case, and integrating this expression on $\Sigma$ gives \eqref{EH-GR potentials relation}. Note that for a diffeomorphism the condition $\mathcal{L}_\xi n_\mu\propto n_\mu$ is quite restrictive. It means that the time component of $\xi$ does not depend on the spatial coordinate, i.e. $\partial_a\xi^T=0$ with $\un=-N\rd T$.

\section{Relationship between $\boldsymbol{\theta_\EH}$, $\boldsymbol{\theta_\GR}$ and $\boldsymbol{\theta_\GH}$}
\label{K-appendix}

In \eqref{GR Lagrangian} we have defined the GR Lagrangian to be
\be\la{LGR}
L_\GR[\tilde{g},{n}]\coloneqq\f{1}{2}\eps\Big(\tilde{R}-(\tilde{K}^2-\tilde{K}^{\mu\nu}\tilde{K}_{\mu\nu})\Big),
\ee
where $\tilde{R}=\tilde{R}(\tg)$ and $\tK_{\mu\nu}=\tg_\mu{}^\alpha\tg_\nu{}^\beta\nabla_\alpha n_\beta$. We also recall the definitions
\be 
\eps_\mu=\pa_\mu\ip\eps,\qquad 
\tilde{\eps}=-\hat{n}\ip\eps,\qquad 
\teps_\mu=\pa_\mu\ip\teps=\hat{n}\ip\eps_\mu,
\ee
which imply that $n^\mu \teps_\mu=0$.

In order to evaluate the pre-symplectic potential associated to \eqref{LGR} we use that
\be
\tilde{K}_{\mu\nu}=\f{1}{2}\tg_{\mu}{}^\alpha\tg_{\nu}{}^\beta \CL_{\hat{n}}\tilde{g}_{\alpha \beta}.
\ee
Defining $\tilde\delta n_\mu\coloneqq\tg_\mu{}^\alpha \delta n_\alpha$, we have the variational identity
\be\label{variation induced K}
\tg_{\mu}{}^\alpha\tg_{\nu}{}^\beta\delta\tK_{\alpha \beta}
&=\f{1}{2} \tg_{\mu}{}^\alpha \tg_{\nu}{}^\beta \left(\CL_{\hat{n}}\delta\tilde{g}_{\alpha \beta}+2 {\nabla}_{(\alpha}\delta n_{\beta)} \right)+ 
\tg_{\mu}{}^\rho\delta \tg_\rho{}^\alpha \tg_{\nu}{}^\beta\CL_{\hat{n}} {g}_{\alpha \beta},\cr
&=\f{1}{2} \tg_{\mu}{}^\alpha \tg_{\nu}{}^\beta \CL_{\hat{n}}\delta\tilde{g}_{\alpha \beta}+\tilde{\nabla}_{(\mu}\tilde\delta n_{\nu)}+\tilde{a}_{(\mu}\tilde\delta n_{\nu)}-\tilde K_{\mu\nu }(n^\alpha \delta n_\alpha).
\ee
For the second line of this identity we have used $ \tg_\rho{}^\alpha=\delta_\rho{}^\alpha+n_\rho{}n^\alpha$ and $n_\alpha\nabla_\mu n^\alpha=0$, which imply that
\be
\big(\tg_{\mu}{}^\rho\delta \tg_\rho{}^\alpha\big) \tg_{\nu}{}^\beta\CL_{\hat{n}} {g}_{\alpha \beta}
&=\big(\tg_{\mu}{}^\rho\delta(n^\alpha n_\rho)\big)\tg_{\nu}{}^\beta\CL_{\hat{n}} {g}_{\alpha \beta}\cr
&=\tg^\rho_\mu n^\alpha\delta n_\rho\tg_{\nu}{}^\beta(\nabla_\alpha n_\beta+\nabla_\beta n_\alpha)\cr
&=\tilde \delta n_\mu\tg_{\nu}{}^\beta\tilde{a}_\beta \cr
&=\tilde \delta n_\mu\tilde{a}_\nu\,,
\ee
and
\be
\tg_{\mu}{}^\alpha \tg_{\nu}{}^\beta {\nabla}_{(\alpha}\delta n_{\beta)}&=
\tg_{(\mu}{}^\alpha  {\nabla}_{\nu)}\delta n_{\alpha}\cr
&=\tilde  {\nabla}_{(\mu}\tilde\delta n_{\nu)} -\left( n_{(\mu} \delta n_\alpha\tilde \nabla_{\nu)} n^\alpha +  n^\alpha \delta n_\alpha   \tilde \nabla_{(\nu} n_{\mu)} \right)\cr
&=\tilde  {\nabla}_{(\mu}\tilde\delta n_{\nu)}-\tilde K_{\mu\nu}(n^\alpha \delta n_\alpha)\,,
\ee
where in the last step we have used again the foliation preserving condition $\delta n_\mu\propto n_\mu$. Using this we finally find the potential
\be
\theta^\mu_\GR=-\f{1}{2}n^\mu(\tK\tilde{g}^{\alpha\beta}-\tK^{\alpha\beta})\delta\tilde{g}_{\alpha\beta}+\tilde{\theta}^\mu_\GR,
\ee
where the second term can be evaluated explicitly, but it does not contribute to the pre-symplectic potential as its pull back on a slice normal to $n_\mu$ vanishes. We thus recover the canonical GR potential.

Let us now focus on the boundary Lagrangian in \eqref{EH=GR+K}. We can write the EH Lagrangian as $L_\EH=L_\GR+\rd L_{\EH/\GR}$ where 
\be
L_{\EH/\GR}=\eps_\mu(n^\mu\tK-\tilde{a}^\mu)=-\teps\tK-\eps_\mu\tilde{a}^\mu
\ee
is the boundary Lagrangian. We would like to evaluate explicitly the variation
\be
\delta L_{\EH/\GR}=-\delta(\teps\tK)-\delta(\eps_\mu\tilde{a}^\mu).
\ee
Using
\be
\delta\eps_\mu=\f{\delta\epsilon}{\epsilon}\epsilon_\mu,
\ee
we get
\be
\delta(\eps_\mu\tilde{a}^\mu)
&=\delta\eps_\mu\tilde{a}^\mu+\eps_\mu\delta\tilde{a}^\mu\cr
&=\delta\eps_\mu\tilde{a}^\mu+\eps_\mu\left(\tilde\delta\tilde{a}^\mu-n^\mu(n_\alpha\delta\tilde{a}^\alpha)\right)\cr
&=-\teps\,\tilde{a}^\mu\delta n_\mu+\epsilon_\mu\left(\tilde\delta\tilde{a}^\mu+\f{\delta \eps}{\eps}\tilde{a}^\mu\right),
\ee
where we have denoted $\tilde{\delta}\tilde{a}^\mu\coloneqq\tg^\mu{}_\alpha\delta\tilde{a}^\alpha$. Using \eqref{mainEH-GR}, we finally obtain the identity
\be\label{relattheta}
\delta L_{\EH/\GR}
&=-\left(\delta(\tilde{\eps}\tilde{K})-\teps\,\tilde{a}^\mu\delta n_\mu\right)-\eps_\mu\left(\tilde\delta\tilde{a}^\mu+\f{\delta\eps}{\eps}\tilde{a}^\mu \right)\cr
&=\tilde{\eps}\,n_\mu\theta^\mu_\EH-\f{1}{2}\tilde{P}^{\mu\nu} \delta\tilde{g}_{\mu\nu}-\tilde{\eps}\,\tilde{\nabla}_\mu\delta n^\mu_\per-\left(\tilde{\delta}\tilde{a}^\mu  +\f{\delta \eps}{\eps}\tilde{a}^\mu \right)\epsilon_\mu\cr
&=(\theta_{\EH}-\theta_{\GR})-\tilde\rd(\delta n^\mu_\perp\teps_\mu)-\delta(\eps\tilde{a}^\mu)\frac{(\eps_\mu + n_\mu \eps)}{\eps},
\ee
where we have denoted 
\be
\theta_\EH\coloneqq\tilde{\eps}\,n_\mu\theta^\mu_\EH,\qquad 
\theta_\GR\coloneqq\f{1}{2}\tilde{P}^{\mu\nu} \delta\tilde{g}^{\mu\nu},\qquad
\tilde{\rd} =\rd x^\alpha \tg_\alpha{}^\beta \pa_\beta.
\ee
The last term in \eqref{relattheta} vanishes when pulled back on a slice normal to $n_\mu$
and the second term can be integrated by part. Integrating this relation therefore gives
\be
\Theta_{\EH/\GR}\coloneqq\int_{\Sigma}(\theta_\EH-\theta_\GR)-\delta\left(\int_\Sigma L_{\EH/\GR}\right)=\int_S\teps_\mu\delta n^\mu_\perp.
\ee
We see that the difference between the pre-symplectic potentials associated to $L_\EH$ and $L_\GR$  is \emph{not} simply given by the total variation $\delta L_{\EH/\GR}$ as often wrongly postulated \cite{Wald:1999wa}. It is given by the combination $\delta L_{\EH/\GR}-\tilde{\rd}\theta_{\EH/\GR}$ where 
\be 
\theta_{\EH/\GR}\coloneqq\teps_\mu\delta n^\mu_\perp,\qquad 
\delta n^\mu_\perp=\f{1}{2}(\delta n^\mu+g^{\mu\alpha}\delta n_\alpha),
\qquad
\teps_\mu=n\ip\pa_\mu\ip\eps.
\ee
The corner pre-symplectic potential can in fact be interpreted as the pre-symplectic potential of the boundary Lagrangian. Indeed, we can express, in agreement with the general theory developed in Appendix \ref{Noethera}, our relationship as 
\be\label{Cmain}
\delta L_{\EH/\GR}= \theta_{\EH}-\theta_{\GR}-\rd\theta_{\EH/\GR}.
\ee

This description is analogous to the bulk variation $\delta L = E_L+\rd \theta_L$. This means that  $  \theta_{\EH} -\theta_{\GR}$ can be interpreted as 
the boundary equation of motion while the relative potential $\theta_{\EH/\GR}$ is (minus) the corner pre-symplectic potential. This is in agreement with \cite{Harlow:2019yfa}.
Our sign conventions are such that $\theta_{\EH/\GR} =-\theta_{L_{\EH/\GR}}$.

Finally, let us proceed to a useful rewriting of the corner pre-symplectic potential. When integrated over a surface one gets that 
\be 
\Theta_{\EH/\GR}=\int_S\theta_{\EH/\GR}=\int_S\teps_\mu\delta n^\mu_\perp=-\int_S\bar{\tilde{\eps}}\,\tilde{s}_\mu\delta n^\mu_\perp,\qquad
\bar{\tilde{\eps}}\coloneqq-\hat{\tilde{s}}\ip\hat{n}\ip\eps,
\ee
where $\tilde{s}^\mu$ is a vector normal to $S$ and to $n_\mu$. The modulus of $\bar{\tilde{\eps}}$ is equal to $\sqrt{q}$:
\be
|\bar{\tilde{\eps}}|=\sqrt{q}. 
\ee
We can write this expression in terms of the normal basis $(\hat{n},\hat{s})$. If one assumes that both $n_\mu$ and $s_\mu$ can be chosen to be normal to $S$, and if one defines the boost angle 
\be
\hat{n}\cdot\hat{s} = \sinh\eta, 
\ee
then we can express the vector $\hat{\tilde{s}}$ as 
\be
\tilde{s}_\mu=\f{s_\mu+n_\mu\sinh\eta}{\cosh\eta}. 
\ee
Hence, using that $(\delta n_\perp^\mu)n_\mu=0$, we get that 
\be 
\Theta_{\EH/\GR} = -\frac12 
\int_S {(\delta n^\mu {s}_\mu + \delta n_\mu s^\mu)} \frac{ \bar{\tilde{\eps}}}{{\cosh\eta}}.
\ee
We can now easily evaluate in the same manner $\Theta_{\EH/\GH}$, namely
\be 
\Theta_{\EH/\GH} = -\frac12 
\int_S {(\delta s^\mu {n}_\mu + \delta s_\mu n^\mu)} \frac{ \tilde{\bar{\eps}}}{{\cosh\eta}}.
\ee
If one uses that
\be
\tilde{\bar{\eps}}=-\bar{\tilde{\eps}}=\f{\hat{s}\ip\hat{n}\ip\eps}{\cosh\eta},
\ee
we get that the relative potential is simply given by 
\be 
\Theta_{\GH/\GR} =\Theta_{\EH/\GR}-\Theta_{\EH/\GH}=\int_S\tilde{\bar{\eps}}\,\delta\eta.
\ee
This means that the relative charge is 
\be
\CH_{\GH/\GR}^S[\xi]=\int_S\tilde{\bar{\eps}}\,\CL_\xi\eta.
\ee
It vanishes if and only if the boost angle is constant on the sphere.


\section{Variation of the diffeomorphism generator}
\label{appendix:Brown-York}

Here we give the proof of \eqref{vector constraint generator}. For simplicity we will do this backwards. We start by writing the total Hamiltonian generator as 
\be
\CH_\GR[\xi]
&=\CH_\GR^\Sigma[\xi]+\CH_\GR^S[\xi] \cr
&=-\int_\Sigma\xi_\nu\tilde{\nabla}_\mu\tilde  P^{\mu\nu}  +\int_\Sigma \tilde\nabla_\mu(\tilde{P}^{\mu\nu}  \xi_\nu)\cr
&=\int_\Sigma\tilde{\nabla}_\mu\xi_\nu \tilde{P}^{\mu\nu}  \cr
&=\f{1}{2}\int_\Sigma\CL_\xi\tilde{g}_{\mu\nu}\tilde{P}^{\mu\nu} .
\ee
Using the Leibniz rule for the Lie derivative and the fact that $\xi$, $\tilde{P}^{\mu\nu}  $ and $\tilde{g}_{\mu\nu}$ are tangent tensors, we have the integrated identity
\be
\int_\Sigma\delta \tilde{g}_{\mu\nu}\CL_\xi\tilde  P^{\mu\nu}  +\tilde{P}^{\mu\nu}  \CL_\xi \delta \tilde{g}_{\mu\nu} 
=\int_\Sigma\CL_\xi(\tilde{P}^{\mu\nu}  \delta \tilde{g}_{\mu\nu}) 
=\int_S\sqrt{q}(\xi^\alpha s_\alpha)\tilde{P}^{\mu\nu} \delta\tilde{g}_{\mu\nu}.
\ee
With the condition $\delta\xi=0$, the variation of the tangent diffeomorphism generator is then given by
\be
\delta\CH_\GR[\xi]
&=\f{1}{2}\int_\Sigma\CL_\xi\tilde{g}_{\mu\nu}\delta \tilde{P}^{\mu\nu} + \CL_\xi( \delta\tilde{g}_{\mu\nu})\tilde{P}^{\mu\nu} \nn\\
&=\f{1}{2}\int_\Sigma\CL_\xi\tilde{g}_{\mu\nu}\delta \tilde{P}^{\mu\nu} -\delta\tilde{g}_{\mu\nu}\CL_\xi \tilde{P}^{\mu\nu} +\CL_\xi(\delta\tilde{g}_{\mu\nu}\tilde{P}^{\mu\nu} )\nn\\
&=-\CL_\xi\ipp\Omega_\GR + \int_S \sqrt{q} (\xi^\alpha s_\alpha)\tilde{P}^{\mu\nu} \tilde{g}_{\mu\nu},
\ee
which gives \eqref{vector constraint generator} when $\xi$ is tangent to the boundary.

\section{$\boldsymbol{2+2}$ decomposition of the Komar charge}
\label{2+2 Komar}

For vector fields which are tangent to both $\Sigma$ and $S$ we can decompose the Komar charge as
\be
\CH[\xi]
&=\f{1}{2}\int_S\sqrt{q}(\tilde{s}^\mu n^\nu-n^\mu\tilde{s}^\nu)\nabla_\mu\xi_\nu\nn\\
&=\f{1}{2}\int_S\sqrt{q}\big(\tilde{s}^\mu\nabla_\mu(n^\nu\xi_\nu)-n^\mu\nabla_\mu(\tilde{s}^\nu\xi_\nu)-\xi_\nu(\tilde{s}^\mu\nabla_\mu n^\nu-n^\mu\nabla_\mu\tilde{s}^\nu)\big)\nn\\
&=\f{1}{2}\int_S\sqrt{q}\big(\tilde{s}^\mu\nabla_\mu(n^\nu\xi_\nu)-n^\mu\nabla_\mu(\tilde{s}^\nu\xi_\nu)-\xi_\nu[\tilde{s},n]^\nu\big)\nn\\
&=\f{1}{2}\int_S\sqrt{q}\big(\eps^{AB}\nabla_{(k_A)}(k_B\cdot\xi_\per)-\xi_\para^a[\tilde{s},n]^bq_{ab}\big),
\ee
with $k_A=k_A^\mu\partial_\mu=(k_0,k_1)=(\tilde{s},n)$ and $\eps^{01}=1$.

\begin{figure}[h]
\begin{center}
\includegraphics{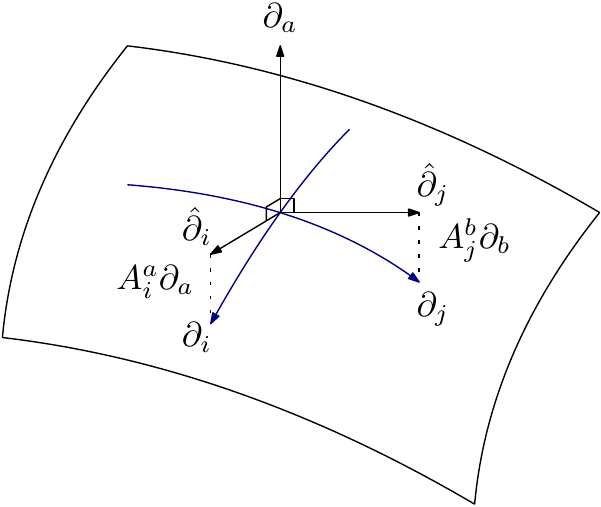}
\end{center}
\caption{Local geometry of the $2+(d-1)$ decomposition.}
\end{figure}

In the $2+(d-1)$ decomposition, we have
\be
g^{\mu\nu}\partial_\mu\partial_\nu=h^{ij}\hat{\partial}_i\hat{\partial}_j+q^{ab}\partial_a\partial_b=h^{ij}(\partial_i+A^a_i\partial_a)(\partial_j+A^b_j\partial_b)+q^{ab}\partial_a\partial_b,
\ee
and
\be
g_{\mu\nu}\rd x^\mu\rd x^\nu=h_{ij}\rd x^i\rd x^j+q_{ab}(\rd y^a-A^a_i\rd x^i)(\rd y^b-A^b_j\rd x^j),
\ee
where $q_{ab}$ is the induced metric on $S$, the normal connection $A^a_i$ is a generalized shift, and the generalized lapse $h_{ij}$ is a $2\times2$ matrix of scalars given by
\be
h_{ij}=k^A_ik^B_j\eta_{AB},
\ee
with $\eta_{AB}=\text{diag}(-1,1)$ a flat 2-dimensional normal metric. The dyad basis of vectors $k_A$ and their dual vector fields $k^A$ are given by
\be
k^A=k^A_i\rd x^i,\q k_A=k_A^i\hat{\partial}_i=k_A^i(\partial_i+A^a_i\partial_a),\q n^i_A=h^{ij}\eta_{AB}k^B_j,\q k_i^A=h_{ij}\eta^{AB}k_B^j.
\ee
With this we indeed have that $k_A\ip k^B=\delta^B_A$ and $k_A\ip(\rd y^a-A^a_i\rd x^i)=0$. With this we now get that
\be
\nabla_{(k_A)}(k_B\cdot\xi_\per)
&=k_A^i(\partial_i+A^a_i\partial_a)(\xi^j_\per k_{Bj})\nn\\
&\stackrel{S}{=}k_A^ik_{Bj}(\partial_i+A^a_i\partial_a)\xi^j_\per\nn\\
&\stackrel{S}{=}h_{jk}k_A^ik_B^k(\partial_i+A^a_i\partial_a)\xi^j_\per\nn\\
&\stackrel{S}{=}h_{jk}k_A^ik_B^k\partial_i\xi^j_\per,
\ee
where we have used the fact that we are considering vector fields such that $\xi_\per\stackrel{S}{=}0$. This then leads to
\be
\eps^{AB}\nabla_{(k_A)}(k_B\cdot\xi_\per)\stackrel{S}{=}\det(k^i_A)h_{jk}\eps^{ik}\partial_i\xi^j_\per\stackrel{S}{=}-\det(k^i_A)h_{jk}\eps^{ki}\partial_i\xi^j_\per\stackrel{S}{=}-\f{1}{\sqrt{q}}Q_j^{~i}\partial_i\xi^j_\per,
\ee
where we have defined the densitized metric
\be
Q_j^{~i}=\sqrt{q}\det(n^i_A)h_{jk}\eps^{ki}.
\ee
Since $\det(h_{jk})=-\det(n^A_i)^2$ and $\det(\eps^{ki})=1$, we get
\be
\det(Q_j^{~i})=-q,\q\tr(Q_j^{~i})=0.
\ee
We also have that
\be
[\tilde{s},n]
&=[k_0,k_1]\nn\\
&=k^i_0k^j_1[\hat{\partial}_i,\hat{\partial}_j]\nn\\
&=\f{1}{2}\eps^{AB}k_A^ik_B^j[\hat{\partial}_i,\hat{\partial}_j]\nn\\
&=\f{1}{2}\det(k_A^i)\eps^{ij}[\hat{\partial}_i,\hat{\partial}_j]\nn\\
&=\det(k_A^i)[\hat{\partial}_0,\hat{\partial}_1]\nn\\
&=\det(k_A^i)[\partial_0+A_0,\partial_1+A_1]\nn\\
&=\det(k_A^i)(\partial_0A_1-\partial_1A_0+[A_0,A_1])\nn\\
&=\f{1}{\sqrt{q}}F,
\ee
where $F$ is the curvature of the connection $A$. Putting this together, we finally get that the Komar charge for a tangential vector field can be rewritten as
\be
\CH[\xi]=-\f{1}{2}\int_SQ_j^{~i}\partial_i\xi^j_\per+\xi_\para^aF_a.
\ee
The normal component of $\CH[\xi,\zeta]$ is therefore
\be
\partial_i[\xi,\zeta]^j
&=\partial_i(\xi^\mu\partial_\mu \zeta^j_\per-\zeta^\mu\partial_\mu \xi^j_\per)\nn\\
&=\partial_i(\xi^k_\per\partial_k\zeta^j_\per-\zeta^k_\per\partial_k\xi^j_\per+\xi^a_\para\partial_a\zeta^j_\per-\zeta^a_\para\partial_a\xi^j_\per)\nn\\
&\stackrel{S}{=}\partial_i\xi^k_\per\partial_k\zeta^j_\per-\partial_i\zeta^k_\per\partial_k\xi^j_\per+\xi^a_\para\partial_a\partial_i\zeta^j_\per-\zeta^a_\para\partial_a\partial_i\xi^j_\per,
\ee
while the tangential component is
\be
[\xi,\zeta]^a
&=\xi^\mu\partial_\mu \zeta^a_\per-\zeta^\mu\partial_\mu \xi^a_\per\nn\\
&\stackrel{S}{=}\xi^b_\para\partial_b\zeta^a_\per-\zeta^b_\para\partial_b\xi^a_\per.
\ee
From this we get
\be
\{Q_i^{~j},Q_k^{~l}\}=\delta^l_iQ_k^{~j}-\delta^j_kQ_i^{~l}\q\{F_a,F_b\}=F_a\partial_b-F_b\partial_a\q\{Q_i^{~j},F_a\}=Q_i^{~j}\partial_a,
\ee
which shows that $F$ generates tangential diffeomorphisms, while $Q$ generates an $\mathfrak{sl}(2,\mathbb{R})^S$ algebra.

\bibliographystyle{bib-style2}
\bibliography{Biblio}

\end{document}

%% file: Macro.tex
\usepackage{caption}

\newcommand{\la}{\label}
\newcommand{\bee}{\nopagebreak[3]\begin{equation*}}
\newcommand{\be}{\begin{equation}}
\newcommand{\ee}{\end{equation}}

\newcommand{\eee}{\end{equation*}}
\newcommand{\beq}{\begin{eqnarray}}
\newcommand{\eeq}{\end{eqnarray}}
\newcommand{\bea}{\begin{eqnarray}}
\newcommand{\eea}{\end{eqnarray}}
\newcommand{\baa}{\nopagebreak[3]\begin{eqnarray*}}
\newcommand{\eaa}{\end{eqnarray*}}

\def\be#1\ee{\begin{align}#1\end{align}}
\def\nn{\nonumber}
\def\q{\qquad}
\def\f{\frac}
\def\eps{\epsilon}
\def\teps{\tilde{\epsilon}}

\def\ip{\lrcorner\,}
\def\pa{\partial}
\def\ipp{\ip}

\def\as{{\ast}}
\def\per{\text{\tiny{$\perp$}}}
\def\para{\text{\tiny{$\parallel$}}}

\def\tr{\mathrm{tr}}

\def\rd{\mathrm{d}}

\def\un{\underline{n}}
\def\us{\underline{s}}

\def\tg{\tilde{g}}

\newcommand{\R}{\mathbb{R}}
\newcommand{\RR}{\mathbb{R}}

\newcommand{\Z}{\mathbb{Z}}

\def\BF{\mathrm{BF}}
\def\EH{\mathrm{EH}}

\def\ECH{\mathrm{ECH}}
\def\GR{\mathrm{GR}}
\def\GH{\mathrm{GH}}

\def\BF{\mathrm{BF}}

\def\tE{\tilde{E}}

\def\tK{\tilde{K}}

\def\CF{\mathcal{F}}
\def\CG{\mathcal{G}}
\def\CH{\mathcal{H}}

\def\CL{\mathcal{L}}

\def\SU{\text{SU}}

\def\SL{\text{SL}}
\def\su{\mathfrak{su}}

\def\sll{\mathfrak{sl}}

\def\te{\tilde{e}}

\def\bmu{{\boldsymbol\mu}}
\def\bnu{{\boldsymbol\nu}}

\numberwithin{equation}{section}